# Ultrafast microwave sensing and automatic recognition of dynamic objects in open world using programmable surface plasmonic neural networks

Qian Ma[1,#,*], Ze Gu[1,#], Zi Rui Feng[1,#], Qian Wen Wu[1], Yu Ming Ning[1], Zhi Qiao Han[1], Rui Si Li[1], Xinxin Gao[1,*] and Tie Jun Cui[1,*]

[1] State Key Laboratory of Millimeter Waves and Institute of Electromagnetic Space, Southeast University, Nanjing, China.
[#] These authors contributed equally: Qian Ma, Ze Gu, Zi Rui Feng
*E-mail: maqian@seu.edu.cn; xxgao@seu.edu.cn; tjcui@seu.edu.cn

**Abstract**

The evolution toward next-generation intelligent sensing requires microwave systems to move beyond static detection and achieve high-speed and adaptive perception of dynamic scenes. However, the existing microwave sensing systems have bottlenecks owing to their sequential digital processing chain, limiting the refresh rates to hundreds of hertz, while the existing integrated microwave processors are lack of programmable and scalable capabilities for robust and open-world deployment. To break the bottlenecks, here we report a programmable surface plasmonic neural network (P-SPNN) that enables real-time microwave sensing and automatic recognition of dynamic objects in open-world environment. With a perception latency of 25 ns and a refresh rate exceeding 10 kHz, the P-SPNN system operates more than two orders of magnitude faster than the conventional millimeter-wave sensors, while achieving an energy efficiency of 17 TOPS/W. With 288 programmable phase-modulated neurons, we demonstrate real-time and robust classification of persons and cars with 91-97 % accuracy in the open-road scenarios. By further integrating beam-scanning function, P-SPNN enables multi-dimensional spatial-temporal-frequency sensing without the digital preprocessing. These results establish P-SPNN as a programmable, scalable, and low-power platform for high-speed perception tasks in realistic world, with broad implications for autonomous driving, intelligent sensing, and next-generation artificial intelligence hardware.

Recently, as the photolithography technology approaches its physical limits, the improvement of computing power and energy efficiency in the traditional silicon-based chips has markedly decelerated. Against this backdrop, the explosive growth in computational demands driven by artificial intelligence (AI) applications (like large language models) has made the development of high-performance and low-power AI computing hardware an urgent priority. Optical neural networks, implemented either as spatial diffraction systems[1,2] or integrated photonic circuits[3-5], have attracted considerable interest as a promising alternative. By leveraging the large-scale matrix computations (essential for the forward propagation of neural networks) executed via electromagnetic (EM) wave at light speed, such systems can deliver exceptional computational throughput and energy efficiency. Notably, the structural design of neurons in diffraction neural networks (DNNs) closely resembles that of meta-units widely employed across the optical[6-8], terahertz[1,9], and microwave regimes[10,11], wherein the EM wave propagation is engineered by precisely controlling its amplitude and/or phase. A variety of architectures has been developed to enable diverse functions, including MNIST handwritten image classification[6,12], multimode fiber image restoration[13], non-reciprocal analogue computing[14], dual adaptive training[15], and optical generative models[16]. Besides, optical reconfigurable DNNs[17,18] and chip solutions[19-22] have been demonstrated to facilitate the minimization. Although significant progress has been made in laboratory platforms, challenges persist in achieving scalable programmability and real-world deployment. For example, the recurrent DNNs typically require a full-spectrum collection system[3,23], which involves high-bandwidth data acquisition [24] and subsequent digital processing. Moreover, the dataset validations are usually performed offline[24,25], which deviates from the purpose of in-situ deployment in open world. Therefore, the current works still hold a gap to an efficient and practical universal computing platform.

Microwave and millimeter-wave sensing technologies serve as a vital complement to the optical sensing owing to their robustness under low-visibility and adverse weather conditions, and hence have been widely applied in the areas of automotive intelligence, the Internet of Things, security imaging[26], and so on. However, conventional millimeter-wave radar modules suffer from a fundamental refresh-rate limit[27] (typically around hundreds of hertz), as they require sequential operations including the transmit modulation, echo mixing, sampling, and digital signal processing. As comparison, the high-performance AI matrix computing abilities of microwave DNNs can simplify these processing links by directly performing computations during the EM wave propagation. DNN can overcome the constrains of frame rate and enable inherent high-speed perception, positioning it as a compelling candidate the for next-generation intelligent sensing systems. Based on diverse metamaterial structures[28-31], the microwave DNNs exhibit similar performance to that in the optical domain to perform image recognition and classification tasks[11,32-34]. They can directly process the microwave information, enabling different applications such as communication encoding and decoding[10,11], target detection[35], echo sensing[33], and adaptive training[11]. Furthermore, the programmability and nonlinearity of the network architectures can be flexibly designed[10,11]. However, the bulky form[11] and limited programmability[10] of these systems restrict their deployment, failing to achieve the inherent high-refresh-rate advantage of the microwave DNNs in practical open-world sensing.

Here, we propose a programmable surface plasmonic neural network (P-SPNN) for ultrafast microwave sensing and accurate recognition of dynamic objects like gestures, persons, and vehicles. Based on the flexible dispersion characteristics and strong field confinements of spoof surface plasmon polaritons (SSPPs), we construct a fully connected planar diffraction neural network. This modular network with 288 programmable neurons allows for smart adaptation to many applications such as image recognition, beam scanning, and object perception. In indoor scenarios, we experimentally verified a recognition accuracy of 95.86% for four sets of dynamic gestures; while in outdoor road scenarios, the P-SPNN-based vehicle platform demonstrated high-speed recognition of persons, cars, and other targets, achieving 91-97% recognition accuracy with a single detection latency of 25 nanoseconds. To cater to the needs for open-world applications, we develop a multi-dimensional data acquisition method that integrates the spectrum, space, and time information. We design a diffraction network training algorithm by taking into account the time-series acquisition data to achieve efficient target recognition. The modular P-SPNN is not only used to train object perception but also to train electromagnetic beam scanning, significantly improving the perception dimensionality. To capture dynamic multidimensional information, the system employs continuous sampling in the time- and frequency-domain superposition to effectively improve the accuracy of object perception. We believe that the programmable planar architecture, with its excellent scalability, is a key enabler for future compact integrated communication and radar systems.

**Problem description and principle**

In open-world complex scenarios, microwave sensing systems are often required to process multiple dynamic targets and directional information in real time[26]. The conventional sensing radars, however, suffer from high system latency due to the sequential processing chains[27] including down-conversion, filtering, and analog-to-digital conversion, which limit their throughput and ultimately delay downstream digital AI tasks. Therefore, to achieve light-speed and high-throughput processing, the future microwave perception systems must shift the computation into the analog domain, in which diffractive neural networks (DNNs) hold significant potentials. Secondly, such systems require substantial programmability, a scalable array of physical neurons, and high integration density, which necessitate a compact and programmable analog computing platform.

To meet these integrated demands, we propose P-SPNN that harnesses the advantages of diffraction-based electromagnetic computing to achieve high-speed microwave sensing and processing with nanosecond-level latency. To realize such dynamic microwave perception, the system must effectively process information along the temporal dimension, while the temporal features are also incorporated into the training data collection process. To further enrich the information content and enhance the recognition robustness, we introduce multiple frequency excitations, enabling a single electromagnetic target to scatter richer spectral signatures. These design considerations lead to a comprehensive training and recognition system based on the P-SPNN architecture shown in Fig. 1a. In the training stage, to enhance the dimensionality of the network's perception, we use the combined echo generated by simultaneous excitations of three frequency signals as the input diffraction information (see Supplementary Note 3 for the detailed structure of training data). In the recognition stage, the pre-detected target information is processed by P-SPNN, which outputs diffraction data to the detector, achieving the superposition and analog-to-digital conversion of different frequency classifications. Finally, the resulting information at each instant is recorded using an FPGA.

The proposed planar P-SPNN is composed of multiple diffractive and phase-modulated layers, as shown in Fig. 1b. To achieve a fully connected network, the diffractive layer is constructed by cross-cascaded SSPP couplers. To flexibly control the electromagnetic waves of each channel, we employ the reflection-type phase shifter on the phase-modulated layer to reconfigure the network's phase weight. The detailed design of structure size is provided in Supplementary Note 2. The simulated amplitude and phase responses of the phase-modulated unit are shown in Figs. 1c and 1d, respectively. As the capacitance of the integrated varactors varies from 0.05 to 0.25 pF, the amplitude difference at 10.2-11 GHz is approximately 2 dB, and the phase shifting range achieves 330°, providing a precondition for comprehensively updating the network weight. The designed SSPP coupler exhibits consistent transmission and coupling parameters, low reflection, and good isolation at the operating frequency (further details in Supplementary Note 2).

To verify the capacity of P-SPNN in perceiving complex and dynamic targets rapidly, we perform two experiments: 1) dynamic gesture combination, transformation, and perception; and 2) real-world road, pedestrian, and vehicle recognition, as shown in Fig. 1e,f respectively. Both experiments achieve recognition accuracy exceeding 90%. Different dynamic targets (gestures, persons and vehicles) possess rich feature information reflected in the microwave echoes. The information captured by the received antenna arrays is directly processed and then recognized by P-SPNN. For more advanced tasks, the network can not only perceive but also modulate the transmitted signals. Specifically, in the person-car recognition experiment, we trained the other P-SPNN as a beam scanner, achieving 100° beam scanning through rapidly switching the weight of the phase-modulated layer, thereby improving the dimensionality of echo information mining. To demonstrate the functional diversity and versatility of P-SPNN, we present two classic image classification cases, MNIST and Fashion-MNIST, achieving the accuracies of 94% and 83%, respectively. For details, please refer to Supplementary Note 1.

**Dynamic gesture recognition**

Based on the proposed P-SPNN architecture, we firstly verify the dynamic gesture recognition task in experiments, as shown in Fig. 2. Hand gestures encompass rich information on the hand shapes and poses, serving as an important and representative testbed for the microwave and millimeter-wave sensing[36,37]. To accurately capture the fine-grained features, it is essential to exploit the scattering responses across multiple frequencies, as different spectral components interact distinctly with the hand geometry. Thus we introduce a diffractive neural architecture that coherently combines the information in both spatial and frequency domains, enabling the network to learn from multi-spectral diffraction patterns and improve the discrimination of complex hand poses. We construct a training and testing hardware system consisting of a transmitting horn antenna, a receiving antenna array, a P-SPNN, a vector network analyzer (VNA), and a switch matrix (Fig. 2a). The gesture sensing area is located 40 cm from the front of the antenna array. The transmitter (Tx) and receiver (Rx) antennas consist of a rectangular horn and a 16 microstrip antenna array, respectively. For training data acquisition, we use a multi-port S-parameter measurement system combining VNA and switch matrix (see Supplementary Note 3 for the detailed structure) to directly collect the reflected field data, which are generated by different gestures transmitted through the horn. As shown in Fig. 2b, we construct the training dataset by sampling the transmission parameters at three discrete frequencies (10.6 GHz, 10.7 GHz, and 10.8 GHz) collected by VNA (see methods for the detailed process).

Given that the speed of gesture changes in the millisecond order is much slower than the speed of diffraction calculation in the nanosecond order, the entire dynamic gesture process can be viewed as a combination of multiple static gesture frames. To enable the network to capture the temporal evolution inherent in the dynamic gestures, the architecture must incorporate the mechanisms for processing information across distinct time instances. Therefore, to classify the dynamic gesture combinations, we adopt a sliding window approach to combine two sequential collection data into one sample. For each gesture combination, which are defined in Fig. 2e-h, the front and rear collections have a fixed time interval of 5s. This interval is chosen so that the relatively faster transition of the gesture would not affect the classification result. To train the programmable nodes inside P-SPNN, we collect the dynamic echo signals at the receiving antenna nodes as the raw data, including four different recipients with different central positions. Then, P-SPNN is trained in conjunction with the digital linear network, which transforms the intensity of the three output ports from SPNN into the four output nodes. For detailed information on the collection results, please refer to Supplementary Note 3. The confusion matrices for the training and test sets are shown in Figs. 2c and 2d, with the training and test accuracy rates of 99.68% and 95.86%, respectively.

As P-SPNN can process the EM echo signals almost instantaneously, the key to achieve dynamic recognition lies in the joint processing of temporal information. Dual sampling slide along the time dimension is presented in the subfigures of instant waveforms of S-parameters shown in Fig. 2e-h. It forms the combined temporal data for the network analysis to complete the dynamic gesture recognition (see Supplementary Note 3 for visualized intermediate results). Please refer to Supplementary Video 1 for experimental demonstration of the dynamic gesture recognition.

**Person-car recognition in open world**

Person-car recognition is a core perceptual task for autonomous driving. However, achieving high-speed and real-time AI perception in complex open-world environments presents big challenges, as it demands not only low-latency and high-refresh-rate sensing but also a system that is programmable to adapt to dynamic scenes and compact enough for practical deployment. To address these challenges, we design a person-car recognition system in the open-world road scenarios. Based on the EM beam scanning area from the vehicle-mounted antenna, we define three recognition scenarios for the left, central, and right regions, as shown in Figs. 3a-c. The left region corresponds to the oncoming lane in the "keep right" driving rule to determine the presence of oncoming vehicles in this region, in which the corresponding scenarios are listed in the table in Fig. 3a. The central region (Fig. 3b) corresponds to vehicles ahead and pedestrians suddenly crossing the lane, hence both human and vehicle need to be identified simultaneously. The right region corresponds to the sidewalk, where persons may be waiting to cross the road, as illustrated in Fig. 3c. Consequently, the primary objection is to detect their presence. More detailed illustration is provided in Supplementary Note 4. The vehicle used to detect the target is equipped with two P-SPNNs for beamforming at the transmitter and computing at the receiver, named as Tx-SPNN and Rx-SPNN, respectively. As shown in Fig. 3d, Tx-SPNN is mounted on the hood of the vehicle to achieve beam scanning by rapidly switching the network weights; while Rx-SPNN is placed on the roof for recognizing and connecting to 12 receiving antennas (eight on the roof and four on the side) for EM information processing. More details on platform installation are provided in Supplementary Note 5. Figure 3e shows the Tx-SPNN port connection configuration, where the top four input ports are fed simultaneously via a power divider, while the outputs are connected to 32 broadband antennas via coaxial adapters (see Supplementary Note 5 for specific configuration). Figure 3f shows Rx-SPNN, where 12 of the input ports are connected to the receiving antennas and 3 of the output ports are cascaded by detectors for further processing.

To meet the P-SPNN beam scanning requirements for the transmitter, we design 11 beam scanning angles (see Fig. 3g) covering from -50° to +50° at 10° step, which are divided into three regions: left, center, and right. To ensure the reliability of the beam scanning, we firstly conduct experimental tests in a microwave anechoic chamber (see Supplementary Note 6). The far-field measurement results are shown in Fig. 3i. The measured results are consistent with the design objectives, with good overall performance, except for slightly higher sidelobes at large scanning angles. For the Tx SPNN beam scanning network, we also used a gradient descent training algorithm (see Supplementary Note 7). The loss values during the optimization process and the far-field pattern correlation matrix are presented in Figs. 3j and h.

In experimental verification of person-vehicle recognition, we conducted data collection and testing in a real-world road environment. As shown in Fig. 4a, the white car (left of the picture) equipped with Rx SPNNs was driven on an open road to collect the data samples, as described in Fig. 3a-c collaboratively. The framework diagram of the test acquisition system is shown in Fig. 4b. The Tx-SPNN, under FPGA control, performs rapid weight switching to achieve beam scanning, with a switching frequency of approximately 10 kHz. The receiving antenna array is used to collect the reflected beams at different angles and send them to Rx-SPNN for compressed processing. After the detection at sparse output ports, the voltages are sampled by a high-speed ADC, which records the perception data and stacks them into the FPGA for processing. The training data collection and testing for the perception network (Rx SPNN) are based on the aforementioned architecture. It is important to note that each beam switching and perception triggering is synchronously controlled by the trigger signal from the FPGA. For specific control flow, refer to Supplementary Note 8.

Figures 4c-e show the selected scenes and classification results from experimental testing for the left, center, and right scanning regions. The left side of each sub-figure depicts the beam scanning region and the corresponding identified target, while the right side shows an image of a real scene and its corresponding depth camera image. During the actual testing, a depth camera was installed on the test vehicle for simultaneous calibration (scene image and depth information) to ensure that the detected target data were within a certain distance range. The target recognition results in the left, center, and right regions achieve accuracies of 97.63%, 91.54%, and 93.57%, respectively. Based on the calculated confusion matrices, we find that classification errors mostly occur for pedestrians' instances, mainly due to their low equivalent scattering area for probing signal reflection. Therefore, we infer that increasing the antenna number for finer spatial resolution or equivalent gain would be benefit to the system's perceptron ability. It should be noted that the proposed Rx/Tx SPNN architecture is highly suitable for such improvements by expanding the physical network's scale without a major redesign or cost boosting. Additionally, we analyze the dynamic recognition ability for the front-facing vehicle recognition to demonstrate the effectiveness of dynamic testing. Figure 4f shows the recognition result in an instatnce where a front-facing vehicle moves from distance to proximity. The recognition waveform exhibits significant jitter as the vehicle approaches, but stabilizes after it fully approaches. In another validation instance, Figure 4g shows the recognition process for a pedestrian crossing the road (only for the center scanning region). When the person is directly in front of SPNN, the target waveform is stably recognized (the middle sub-figure state). Please refer to Supplementary Videos 2 and 3 for the experimental demonstration of person-car dynamic recognition.

Finally, we illustrate some key metrics of the in-situ recognition system quantitatively in Fig. 5. The latency for one measurement for each scanning pattern includes the propagation time in free space and dual SPNN platforms, together with the detector response time. Based on the oscilloscope waveforms in Fig. 5a, the latency is approximately 25ns in the vehicle recognition. We remark that this time represents the response time for detecting one beam angle. Each complete cycle includes 11 such beam angles, and each round of beam scanning and inspection takes approximately 104 us. More details are given in Supplementary Note 8. The system overhead primarily comes from beam scanning weight switching and subsequent digital processing. However, SPNN, benefiting from the computational speed at nearly the speed of light, represents a minimal overhead in overall system processing cycle, highlighting its enormous computational potential. Therefore, the current system holds a promising prospect when the refreshing speed for beam scanning could be further increased. The subsequent linear transformation could also be performed in analog platforms like the memristor crossbars[38] to further compress the delay, where the processing bandwidth could easily meet the demand for the current systems. In Fig. 5b, we present the power consumption of different modules in the current system. Currently, most energy is consumed on the wideband power amplifier, which is limited by its innate low power added efficiency (PAE). We note that a relative high-power probing source is necessary and common for similar long-range detection tasks. The merit of the proposed analog neural network lies in its instantaneous processing capability and low-power property at the receiving end. In specific, the fabricated Tx and Rx SPNNs draw a power of less than 1W respectively, mainly arising from the multi-channel DACs. Detailed setups and evaluations are provided in Methods and Supplementary Information.

In addition, we perform a series of numerical simulations to prove the effectiveness of the introduced dual SPNNs, in which part of the datasets is shadowed in network training. In Fig. 5c, we demonstrate the diagrams for potential obstacle recognition under specific scenarios when the Rx SPNN and Tx SPNN are excluded from the system. The classification accuracy and loss data are shown in the bar charts of Fig. 5d. When the Rx SPNN is excluded (middle part of Fig. 5c), the perceptron's ability is limited by the sparse receiving antennas without the analog domain preprocessing. When the Tx SPNN is excluded from the system (right part of Fig. 5c), the spatial probing ability is compromised. Compared with the standard dual-SPNN setup, the average classification drops by 7.9% and 14.8% when the Rx and Tx SPNNs are missing in the system, respectively. The results prove that the introduction of SPNN is necessary for the system's perceptron enhancement. Meanwhile, we evaluate the network's performance when reducing the scanning pattern number and receiving output port number in the current network architecture. Based on the accuracy matrix shown in Fig. 5e, we find that the RX port number imposes less influence on the overall performance. It means that the detectors at the output of Rx SPNN could be further optimized, which would further facilitate the processing delay and power in prospect.

## Discussion

We presented a programmable planar SPNN that fundamentally advances microwave sensing by achieving nanosecond-level latency and a system refresh rate exceeding 10 kHz, thereby overcoming the frame-rate bottleneck of the traditional radar-based perception systems. We not only validated it as a modular programmable computing platform for image recognition, but also demonstrated its capability for complex dynamic perception of gestures and person-car targets in open-road environments. We constructed three programmable P-SPNN modules with a total of 288 phase-programmable nodes, achieving accuracies of 94% and 95.86% in the MNIST-image recognition and dynamic gesture recognition, respectively. For the first time, we show that such a high-speed analog neural network can reliably classify the cars and persons in the open-road scenarios, with a single-perception latency of 25 ns. For complex scenes across different sensing areas on open roads, based on experimentally acquired space-time-frequency data, we achieved the scene classification accuracy between 91% and 97%, confirming reliable real-world perception performance. Furthermore, we demonstrated the flexible scalability of the proposed planar architecture. In experiments, the modular P-SPNN not only functions as the hardware for recognizing scattered electromagnetic signals, but also serves as a versatile platform for EM information processing, addressing diverse signal processing requirements in the radar and wireless communications.

It is worth noting that the modular design of the proposed planar programmable network allows flexible expansion. In the future, more integrated designs could incorporate control circuits on the backside of P-SPNN, enabling an ultra-low profile and high-density stacking (see Methods). Additionally, the connection parts on both sides of the SPNN edge can be changed to SMA ports, which can facilitate lateral expansion of the single-layer diffractive network scale (see Methods and Supplementary Note 10). Taking an expanded network as an example, the computational capacity can be increased to 80 TOPS. The main power consumption of P-SPNN is originated from the control module of varactor diodes. Currently, the average power consumption of our multi-channel DAC chip control is about 6 mW/channel, resulting in an overall network energy efficiency of 17TOPS/W. Note that the energy efficiency can be further improved by simplifying the control module and adopting high-performance chips.

## Methods

### Control and detection of P-SPNN

Each programmable neuron (i.e., phase-shifting unit) in the P-SPNN consists of four varactor diodes and is controlled by a bias voltage control line. The control line for each neuron is connected to the output channel of the DAC control board via a control cable on the back of SPNN. In this work, we use an AD5370 chip array for control, employing a modular design. In the static modulation, each DAC module is programmed and synchronously controlled via the integrated serial port of MCU (STM32F103C8T6). In the detection section, we use a high-speed detector chip (LCT5564) in conjunction with an ADC chip (AD7616) for the real-time sampling. The detector response time is approximately 11 ns, and the ADC detection readout time is approximately 0.4 µs. Detailed instructions for the control and acquisition loop are provided in Supplementary Note 8.

**Training data collection**

**(1) Image classification.** In this part, the data was based on the MNIST handwritten digit dataset. Through pixel recombination, the original two-dimensional matrix was transformed into a 1*32 matrix, corresponding to the input (port 1-32) of the P-SPNN network.

**(2) Dynamic hand-gesture classification.** First, we determined the overall installation location of the transceiver antennas and connected them to a four-port vector network (Siglent SNA6144A) via a matrix switch (Siglent SSM5144A) to collect S-parameters. The system architecture is referenced in (Supplementary Note 3). During the data acquisition process, the subjects kept their hands stable in front of the transceiver antenna array and collected data by switching between different hand gestures. Each set of gestures was collected 50 times sequentially, and the different hand gestures were switched in a loop to complete all data acquisition.

**(3) Person-car classification.** During the data collection, we measure the transmission coefficients between the single source and the collected antenna array that are tied to the roof of the measured car. In the subsequent real deployment, this antenna array would directly connect to the SPNN input ports. Here, we use an RF matrix switch and a vector network analyzer to collect the partially S-matrix data. For each sample, the data has a dimension of $11\times12\times3$. The three dimensions represent the beam angles, the number of antennas and three frequency points, respectively.

**Model calibration of P-SPNN**

To obtain the accurate numerical model of the fabricated P-SPNN sample, the cascaded diffraction matrices need to be processed and tested individually in order to extract accurate diffraction matrix parameters. Supplementary Note 9 provides the image and measured results of the fabricated cascaded diffraction matrices without phase-modulation unit. To prevent phase shift errors in the transmission line, we use the shortest possible connection wire to the probe connector. In addition, individual phase shifters were also fabricated and measured to extract the true transmission and reflection coefficients.

**Scale expansion of P-SPNN**

The proposed P-SPNN is a planar structure that can be stacked and expanded using the connection method shown in Fig. S1 (Supplementary Note 1). However, the method in Fig. S1 cannot expand the lateral scale (the scale of neurons per layer), so lateral cascading expansion is required. To achieve this expansion, we need to modify the side ports of the P-SPNN. As shown in Fig. S10 (Supplementary Note 10), some ports need to be externally connected to SMA connectors for inter-board interconnection, while some existing connections need to be disconnected. Finally, a large network consisting of 8 P-SPNNs (768 programmable neurons in total) can be constructed, as shown in Fig. S11(Supplementary Note 10). It should be noted that by using flexible connecting lines, this large network can also significantly compress its volume through a folding method similar to that shown in Fig. S1.

**Acknowledgements**

The work iwas supported by the National Natural Science Foundation of China (62288101 and 62301147), the Natural Science Foundation of Jiangsu Province (BK20230822), the National Key Research and Development Program of China (2022YFA1404903), Jiangsu joint laboratory of multidimensional perceptual information technology (bm2022017), Special Fund for Key Basic Research in Jiangsu Province (BK20243015), the Major Project of Natural Science Foundation of Jiangsu Province (BK20212002, BK20210209), Young Elite Scientists Sponsorship Program by CAST (2022QNRC001), the State Key Laboratory ofMillimeter Waves, Southeast University, China (K201924), the Fundamental Research Funds for the Central Universities (2242023K5002, 2242022R20017), the 111 Project (1112-05), the China Postdoctoral Science Foundation (2021M700761, 2022T150112) , and the Independent Research Fund of the State Key Laboratory of Millimeter Waves (Grant No. Z202503-02, Z202502-03).

**References**

1   Lin, X. *et al.* All-optical machine learning using diffractive deep neural networks. *Science* **361**, 1004-1008 (2018). https://doi.org/10.1126/science.aat8084
2   Luo, Y. *et al.* Design of task-specific optical systems using broadband diffractive neural networks. *Light Sci Appl* **8**, 112 (2019). https://doi.org/10.1038/s41377-019-0223-1
3   Xu, X. *et al.* 11 TOPS photonic convolutional accelerator for optical neural networks. *Nature* **589**, 44-51 (2021). https://doi.org/10.1038/s41586-020-03063-0
4   Shen, Y. *et al.* Deep learning with coherent nanophotonic circuits. *Nature Photonics* **11**, 441-446 (2017). https://doi.org/10.1038/nphoton.2017.93
5   Feldmann, J., Youngblood, N., Wright, C. D., Bhaskaran, H. & Pernice, W. H. P. All-optical spiking neurosynaptic networks with self-learning capabilities. *Nature* **569**, 208-214 (2019). https://doi.org/10.1038/s41586-019-1157-8
6   Luo, X. *et al.* Metasurface-enabled on-chip multiplexed diffractive neural networks in the visible. *Light Sci Appl* **11**, 158 (2022). https://doi.org/10.1038/s41377-022-00844-2
7   Goi, E., Schoenhardt, S. & Gu, M. Direct retrieval of Zernike-based pupil functions using integrated diffractive deep neural networks. *Nat Commun* **13**, 7531 (2022).

https://doi.org/10.1038/s41467-022-35349-4

8   Wang, T. *et al.* Image sensing with multilayer nonlinear optical neural networks. *Nature Photonics* **17**, 408-415 (2023). https://doi.org/10.1038/s41566-023-01170-8

9   Gao, X. X. *et al.* Terahertz spoof plasmonic neural network for diffractive information recognition and processing. *Nature Communications* **15**, 6686 (2024). https://doi.org/ARTN 668610.1038/s41467-024-51210-2

10  Gao, X. X. *et al.* Programmable surface plasmonic neural networks for microwave detection and processing. *Nature Electronics* **6**, 319-328 (2023). https://doi.org/10.1038/s41928-023-00951-x

11  Liu, C. *et al.* A programmable diffractive deep neural network based on a digital-coding metasurface array. *Nature Electronics* **5**, 113-122 (2022). https://doi.org/10.1038/s41928-022-00719-9

12  Qu, G. *et al.* All-Dielectric Metasurface Empowered Optical-Electronic Hybrid Neural Networks. *Laser & Photonics Reviews* **16**, 2100732 (2022). https://doi.org/https://doi.org/10.1002/lpor.202100732

13  Yu, H. *et al.* All-optical image transportation through a multimode fibre using a miniaturized diffractive neural network on the distal facet. *Nature Photonics* (2025). https://doi.org/10.1038/s41566-025-01621-4

14  Li, X. *et al.* Nonreciprocal surface plasmonic neural network for decoupled bidirectional analogue computing. *Nature Communications* **16** (2025). https://doi.org/10.1038/s41467-025-63103-z

15  Zheng, Z. *et al.* Dual adaptive training of photonic neural networks. *Nature Machine Intelligence* **5**, 1119-1129 (2023). https://doi.org/10.1038/s42256-023-00723-4

16  Chen, S., Li, Y., Wang, Y., Chen, H. & Ozcan, A. Optical generative models. *Nature* **644**, 903-911 (2025). https://doi.org/10.1038/s41586-025-09446-5

17  Fu, P. *et al.* Reconfigurable metamaterial processing units that solve arbitrary linear calculus equations. *Nat Commun* **15**, 6258 (2024). https://doi.org/10.1038/s41467-024-50483-x

18  Zhou, T. *et al.* Large-scale neuromorphic optoelectronic computing with a reconfigurable diffractive processing unit. *Nature Photonics* **15**, 367-373 (2021). https://doi.org/10.1038/s41566-021-00796-w

19  Chen, Y. *et al.* All-analog photoelectronic chip for high-speed vision tasks. *Nature* **623**, 48-57 (2023). https://doi.org/10.1038/s41586-023-06558-8

20  Xue, Z. *et al.* Fully forward mode training for optical neural networks. *Nature* **632**, 280-286 (2024). https://doi.org/10.1038/s41586-024-07687-4

21  Xu, Z. *et al.* Large-scale photonic chiplet Taichi empowers 160-TOPS/W artificial general intelligence. *Science* **384**, 202-209 (2024). https://doi.org/10.1126/science.adl1203

22  Wang, C., Cheng, Y., Xu, Z., Dai, Q. & Fang, L. Diffractive tensorized unit for million-TOPS general-purpose computing. *Nature Photonics* (2025). https://doi.org/10.1038/s41566-025-01749-3

23  Ji, J. *et al.* Synthetic-domain computing and neural networks using lithium niobate integrated nonlinear phononics. *Nature Electronics* **8**, 698-708 (2025). https://doi.org/10.1038/s41928-025-01436-9

24  Govind, B., Anderson, M. G., Wu, F. O., McMahon, P. L. & Apsel, A. An integrated microwave neural network for broadband computation and communication. *Nature Electronics* **8**, 738-750 (2025). https://doi.org/10.1038/s41928-025-01422-1

25  Davis, R., Chen, Z., Hamerly, R. & Englund, D. RF-photonic deep learning processor with Shannon-limited data movement. *Science Advances* **11**, eadt3558 (2025). https://doi.org/doi:10.1126/sciadv.adt3558

26  Kong, H., Huang, C., Yu, J. D. & Shen, X. M. A Survey of mmWave Radar-Based Sensing in Autonomous Vehicles, Smart Homes and Industry. *Ieee Communications Surveys and Tutorials* **27**, 463-508 (2025). https://doi.org/10.1109/comst.2024.3409556

27  Hakobyan, G. & Yang, B. High-performance automotive radar: A review of signal processing algorithms and modulation schemes. *IEEE Signal Processing Magazine* **36**, 32-44 (2019).

28  Chen, L. *et al.* Space-Energy Digital-Coding Metasurface Based on an Active Amplifier. *Physical Review Applied* **11**, 6 (2019). https://doi.org/10.1103/PhysRevApplied.11.054051

29  Gao, X. *et al.* Programmable Hybrid Circuit Based on Reconfigurable SPP and Spatial Waveguide Modes. *Advanced Materials Technologies* **5** (2019). https://doi.org/10.1002/admt.201900828

30  Ma, Q. *et al.* Controllable and Programmable Nonreciprocity Based on Detachable Digital Coding Metasurface. *Advanced Optical Materials* **7**, 1901285 (2019). https://doi.org/ARTN 190128510.1002/adom.201901285

31  Ning, Y. M., Ma, Q., Xiao, Q., Gu, Z. & Cui, T. J. Reprogrammable Nonlinear Transmission Controls Using an Information Metasurface. *Advanced Optical Materials* **12** (2024). https://doi.org/10.1002/adom.202301525

32  Gu, Z., Ma, Q., Gao, X. X., You, J. W. & Cui, T. J. Classification of Metal Handwritten Digits Based on Microwave Diffractive Deep Neural Network. *Advanced Optical Materials* **12** (2024). https://doi.org/10.1002/adom.202301938

33  Gu, Z., Ma, Q., Gao, X. X., You, J. W. & Cui, T. J. Direct electromagnetic information processing with planar diffractive neural network. *Science Advances* **10**, eado3937 (2024). https://doi.org/ARTN eado393710.1126/sciadv.ado3937

34  Ning, Y. M. *et al.* Mechanically Programmable Diffractive Neural Networks Based on Pancharatnam-Berry Phase. *Advanced Functional Materials* (2025). https://doi.org/10.1002/adfm.202512689

35  Ning, Y. M. *et al.* Multilayer nonlinear diffraction neural networks with programmable and fast ReLU activation function. *Nature Communications* **16** (2025). https://doi.org/10.1038/s41467-025-65275-0

36  Li, L. *et al.* Intelligent metasurface imager and recognizer. *Light: Science & Applications* **8**, 1-9 (2019). https://doi.org/10.1038/s41377-019-0209-z

37  Ma, Q. *et al.* Intelligent Hand-Gesture Recognition Based on Programmable Topological Metasurfaces. *Advanced Functional Materials* **35**, 2411667 (2025). https://doi.org/10.1002/adfm.202411667

38  Keshavarz, R., Zelaya, K., Shariati, N. & Miri, M.-A. Programmable circuits for analog matrix computations. *Nature Communications* **16**, 8514 (2025). https://doi.org/10.1038/s41467-025-63486-z

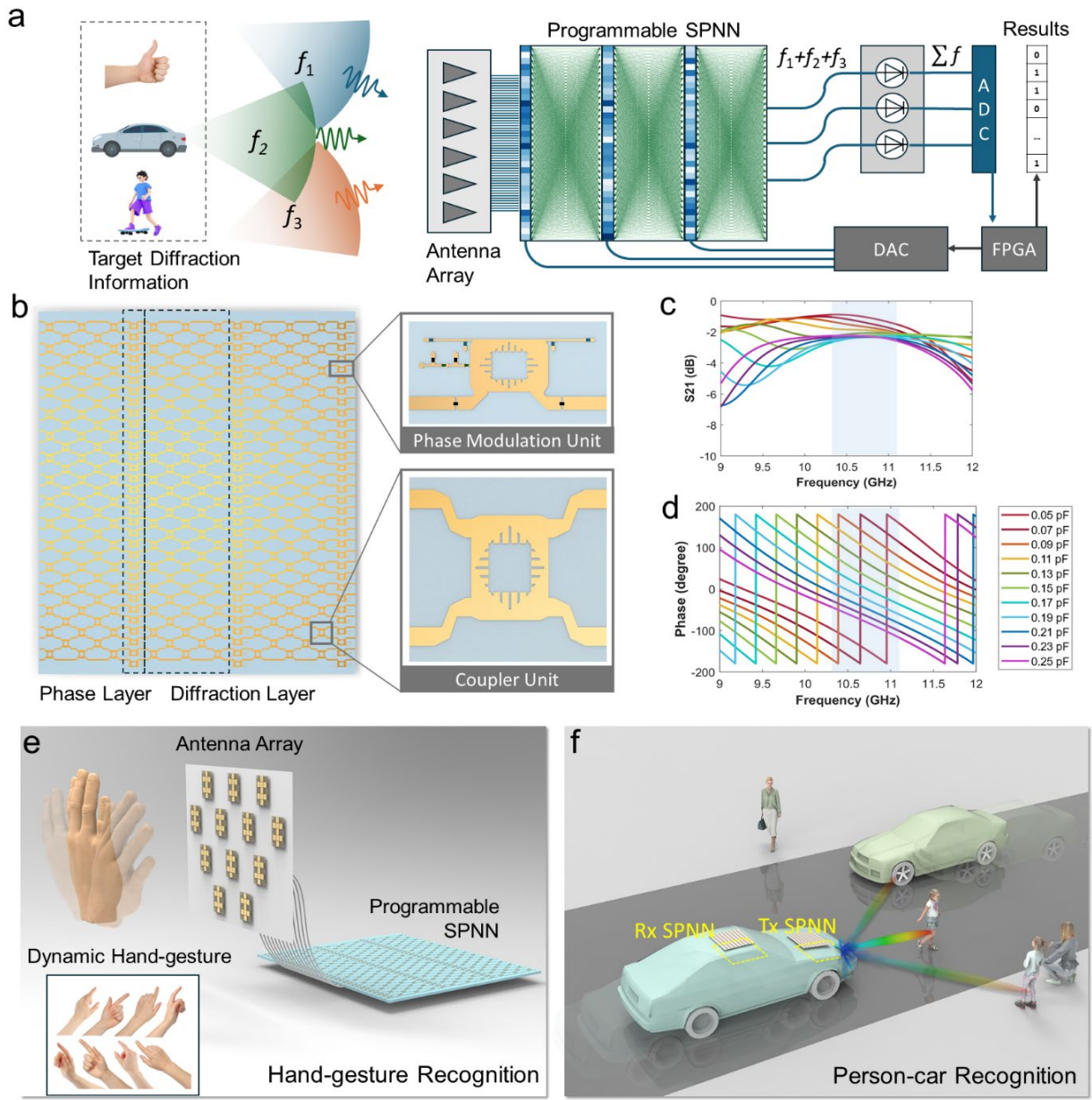

**Fig. 1 Schematic of the programmable surface plasmonic neural networks (P-SPNN) for dynamic recognition in open world. a,** Architecture of the target data acquisition and testing system based on P-SPNN. The echo signals at three frequencies are firstly collected by the front-end antenna array to construct multi-frequency and multi-dimensional training data. The P-SPNN then performs frequency superposition calculations to achieve the target recognition. **b,** P-SPNN structure diagram. A single P-SPNN module consists of three phase layers and three diffraction layers. The phase layer is composed of programmable phase modulation units, and the diffraction layer is composed of cascaded couplers. **c,** Transmission amplitudes of the programmable phase shifter at different phase shifts. **d,** Programmable phase responses at different phase shift states. **e,** The schematic of the hand-gesture recognition. By capturing hand-scattered echoes at high speed, P-SPNN can accurately identify the dynamic gestures. **f,** Schematic diagram of the human-vehicle recognition. P-SPNN is installed on a vehicle platform to construct a high-speed perception system that accurately identifies the human and vehicle targets ahead on the open roads.

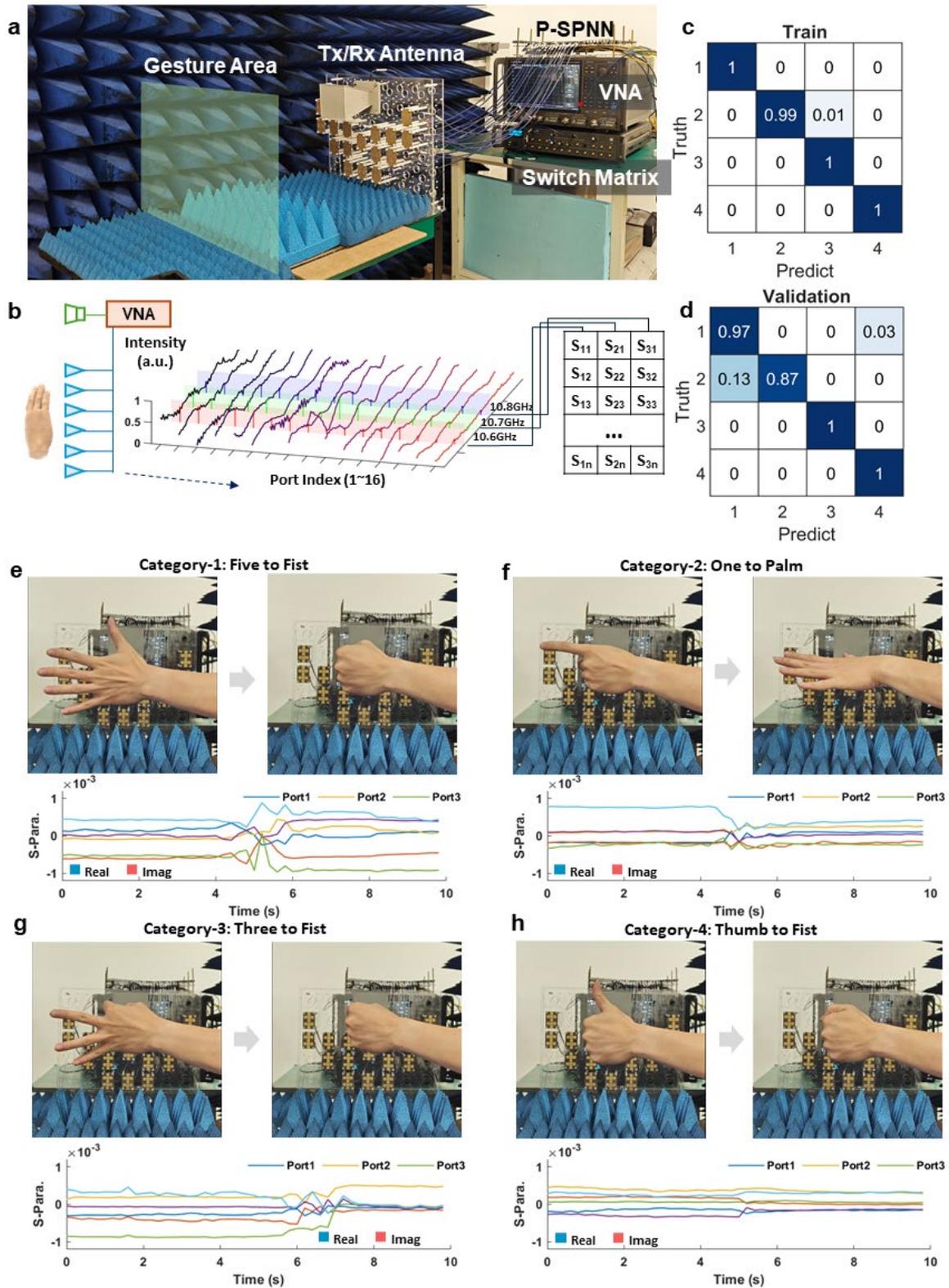

**Fig. 2 Experiment verification of dynamic hand gesture recognition based on P-SPNN. a,** Diagram of the training data collection and experimental testing scenario for the hand gesture recognition. **b,** Flowchart of training/testing data collection and processing. **c-d,** Confusion matrices of the training/ validation samples based on the measured results for four dynamic gestures. **e-h,** Photos of the four dynamic gestures and dynamic waveforms of the SPNN output envelop during real-time testing.

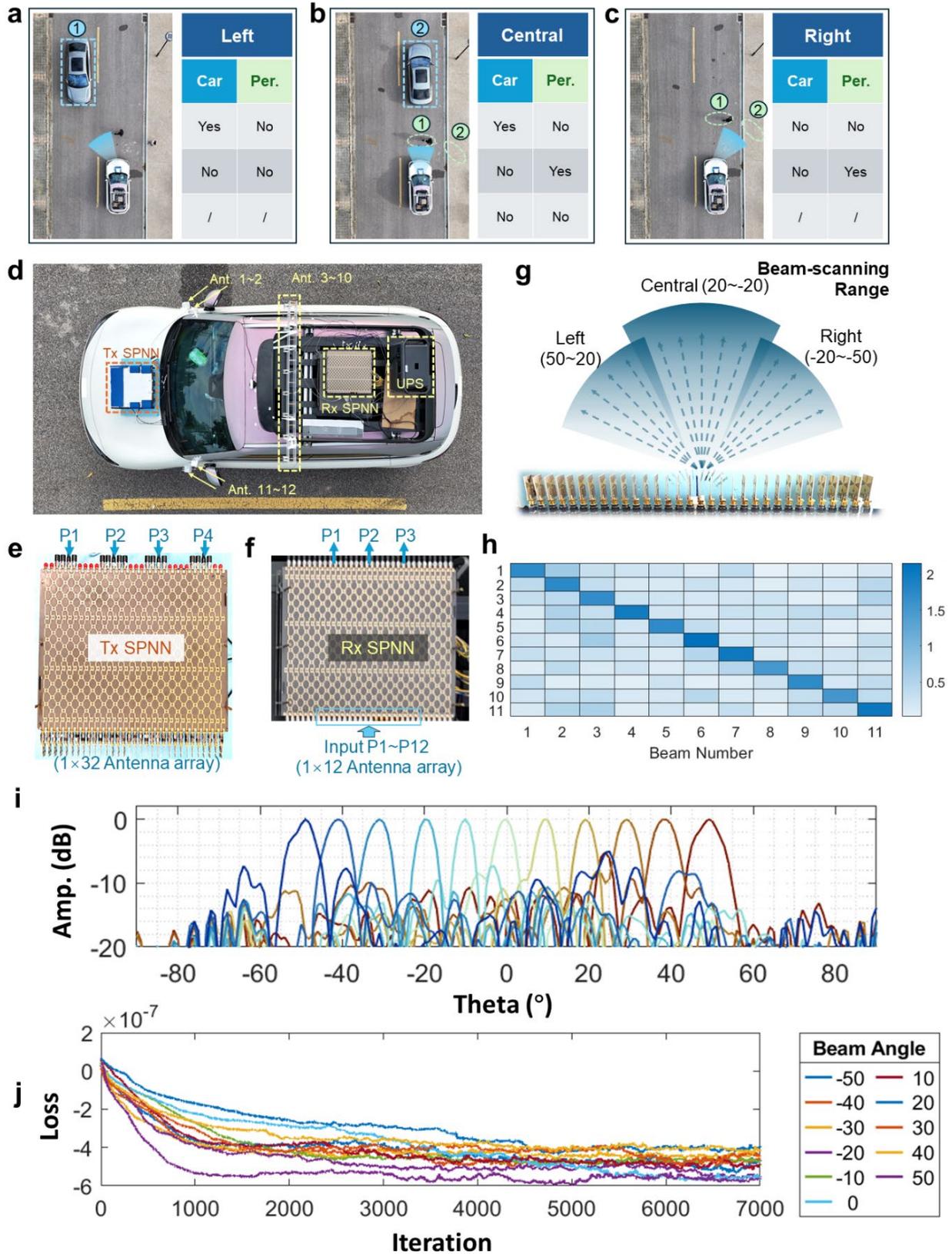

**Fig. 3 Schematic diagram and experimental setup for the human-vehicle recognition. a-c,** Target recognition in the left, center, and right sensing zones for the right lane. The left zone (in **a**) determines the presence of a vehicle, the center zone (in **b**) determines whether a person or vehicle is present, and the right zone (in **c**) determines the presence of a person. **d,** Experimental setup of the vehicle platform.

The Tx SPNN and Rx SPNN are mounted on the front and roof of vehicle, respectively. The receiving sensing antenna array is placed on the side and roof of the vehicle. **e,** Port configuration of the Tx SPNN, where the output ports are connected to a homogeneous 32-element antenna array. The four input ports (P1-P4) are simultaneously excited via a power divider. **f,** Port configuration of the Rx SPNN, where 12 of the input ports are connected to the receiving antennas and 3 of the output ports (P1-P3) are cascaded by detectors for further processing. **g,** Schematic diagram of the Tx SPNN beam scanning. The beam scanning angle is divided into three zones: left, center, and right. **h,** Far-field pattern correlation matrix between the 11 individual beams after the Tx SPNN training is completed. **i,** Far-field radiation patterns at the 11 beam scanning angles tested in experiments. **j,** Loss curve during the Tx SPNN training.

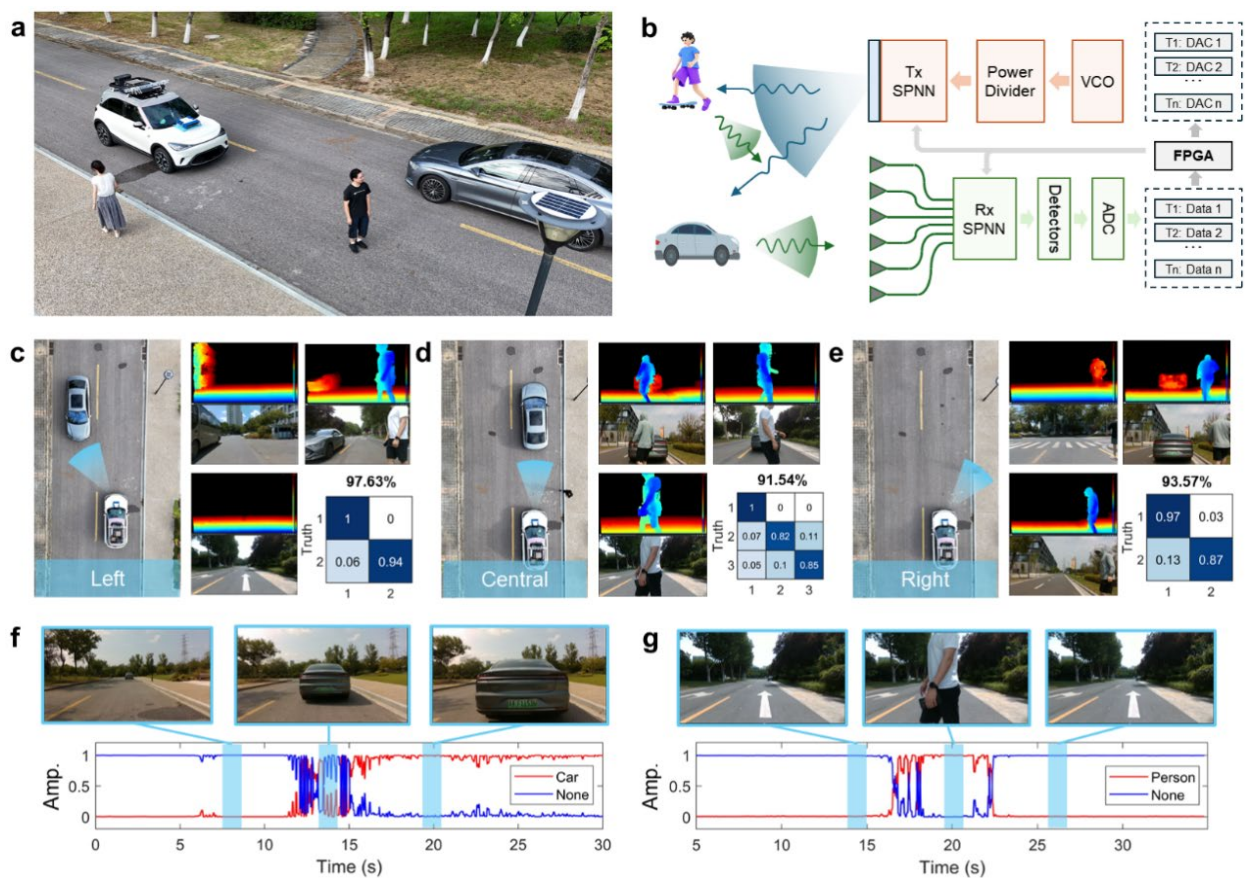

**Fig. 4 Experimental description and test results of the human-vehicle recognition. a,** Experimental scenario for the human-vehicle target recognition in an open world. **b,** P-SPNN-based system framework for the human-vehicle recognition. **c-e,** Experimental results for the left, center, and right perception areas in the right lane. The left sub-image is a schematic of the scene; the middle sub-image includes the calibration data collected by the depth camera for the corresponding scene; the lower right corner shows the confusion matrix for recognition. **f,** Dynamic recognition results for the vehicle ahead during the dynamic driving. As the vehicle approaches, the recognition target gradually stabilizes. **g,** Dynamic recognition results for a pedestrian crossing ahead during the dynamic driving. When the pedestrian appears directly in the forward area, the target is accurately recognized.

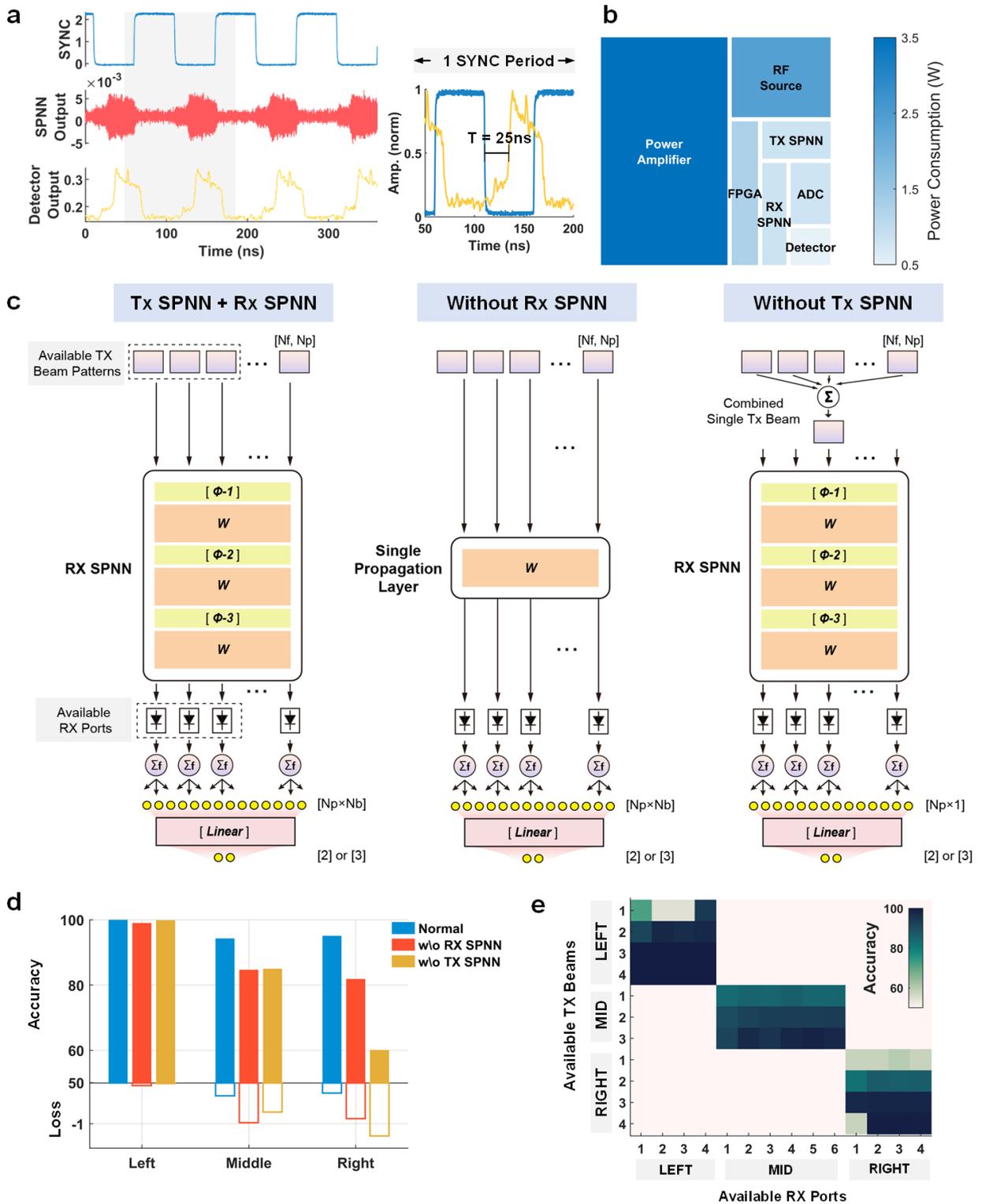

**Fig. 5 Performance analysis of the vehicle recognition using SPNN. a,** Waveforms of (i) synchronize signal of the pulses from the RF signal source, (ii) SPNN output, and (iii) detector envelope output. The processing latency is approximately 25ns, including the propagation delay, and the processing latency of SPNNs and detectors. **b,** The power consumption of different modules in the current system. **c,** The diagrams of the data links in the standard setup and when Rx SPNN and Tx SPNN is removed. **d,** The classification accuracy using different raw datasets. When Tx SPNN is excluded from the system, the probing ability is compromised. When Rx SPNN is excluded, the perceptron's ability is limited by the sparse receiving antennas. **e,** The accuracy for vehicle recognition when using different sets of probing signals and receiving antenna.

Supplementary Information for

# Ultrafast microwave sensing and automatic recognition of dynamic objects in open world using programmable surface plasmonic neural networks

Qian Ma[1,#,*], Ze Gu[1,#], Zi Rui Feng[1,#], Qian Wen Wu[1], Yu Ming Ning[1], Zhi Qiao Han[1], Rui Si Li[1], Xinxin Gao[1,*] and Tie Jun Cui[1,*]

[1] State Key Laboratory of Millimeter Waves and Institute of Electromagnetic Space, Southeast University, Nanjing, China.

[#] These authors contributed equally: Qian Ma, Ze Gu, Zi Rui Feng

*E-mail: maqian@seu.edu.cn; xxgao@seu.edu.cn; tjcui@seu.edu.cn

## This file includes:

**Supplementary Note 1: Experimental verification of P-SPNN on classical image dataset classification**

**Supplementary Note 2: Detailed structure of SPNN**

**Supplementary Note 3: Training and test of dynamic-gesture classification**

**Supplementary Note 4: Description on the classification of person and car**

**Supplementary Note 5: Installation details of the vehicle-mounted platform for person-car recognition**

**Supplementary Note 6: Far-field measurements of Tx SPNN for beam-scanning**

**Supplementary Note 7: Training of person-car classification**

**Supplementary Note 8: Processing time of P-SPNN**

**Supplementary Note 9: Experimental test of the SPNN diffraction layer for S-parameter matrix**

**Supplementary Note 10: Scale expansion of P-SPNN**

**Supplementary Figures S1-S14**

**Supplementary Tables S1**

**References**

# Supplementary Note 1: Experimental verification of P-SPNN on classical image dataset classification

Before the real-scenario deployment, we first demonstrate the classical inference ability of the fabricated SPNN platforms for neural network computing. As shown in the picture of Fig. S1, we construct a fully connected neural network with 9 intermediate layers by cascading three identical SPNN boards (each containing three propagation layers and three modulation layers) using U-shape RF cables. Each column of the modulation layers contains 32 programmable phase shifters that are connected to a customized voltage driver board (See Methods for hardware implementation). Two classic image datasets, i.e., MNIST and fashion-MNIST are chosen as the training datasets. To fit the input dimensions of the SPNN, the original 784-pixel images first go through a two-dimensional discrete Fourier transformation (2D-DFT) and the 32 components that are closest to the DC components are extracted. We note here that this linear mapping could be efficiently implemented in the analog domain using Fourier optics principles in the 3D space. The output ports are distributed evenly across the 32 available output ports. Based on the measured diffraction matrix from Supplementary Note 2, we build the neural network model on computers and execute the gradient-based optimization process. After the algorithm converges, the parameters (phase shifting values) are mapped into the varactors' voltages and transmitted to the corresponding driver boards.

To simulate the all-analog inference process, we directly collect the transmission matrix between the input ports and output ports using a manual port selection process. Then the output responses of the validation dataset samples are acquired using a simulated propagation process performed on a digital computer.

Here, we firstly chose five categories for each dataset and demonstrate the simulated and measured results in Fig. S2. In Fig. S2a, the images of handwritten digit samples for the five categories are shown as examples. The numeric simulation presents an accuracy of 94.5% on the validation dataset while 93.8% on the measured network. The confusion matrixes for each are plotted in Figs. S2b and S2c. For Fashion-MNIST shown in Fig. S2d, the images of fashion-MNIST samples for the five categories are given as examples. The simulated and measured results illustrate an accuracy of 83.4% and 83.1% respectively, as listed in Figs. S2e and S2f. The good alignment between the simulation and experiment validation could be attributed to the in-situ calibration processes (i.e. the remeasurement of the propagation matrix and mapping inside the phase shifter) on the fabricated platforms that minimize the modelling error.

Additionally, we demonstrate the 10-category classification performance of the two datasets. For MNIST images, the 10-category samples are shown in Fig.S3a. For MNIST dataset, the accuracy rate drops to 79.0% (from 80.1% of training set in Fig.S3b) on the validation dataset. For Fashion-MNIST dataset, the 10-category samples are shown in Fig.S3d. The accuracy rates of training and test are 68.86% and 67.9%, as listed in Fig. S3e and S3f. We remark here the performance degradation arises from both a limited trainable variables number (in the current configuration, the number is 32*9=288) and the compressed input information of the images (32 input channels).

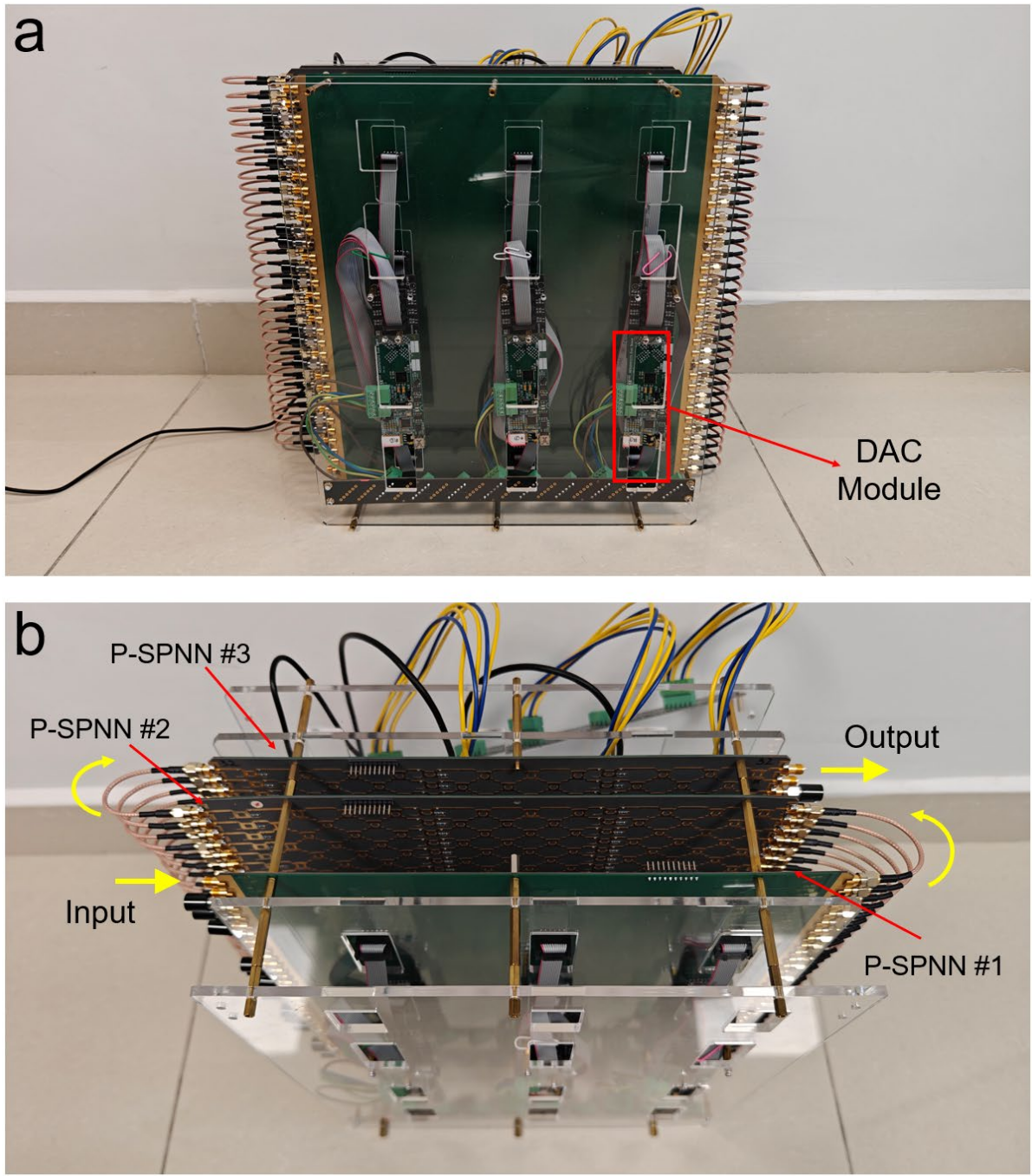

**Fig. S1 The fabricated sample of three cascaded P-SPNNs. a.** Front view of the sample. **b.** Top view of the sample.

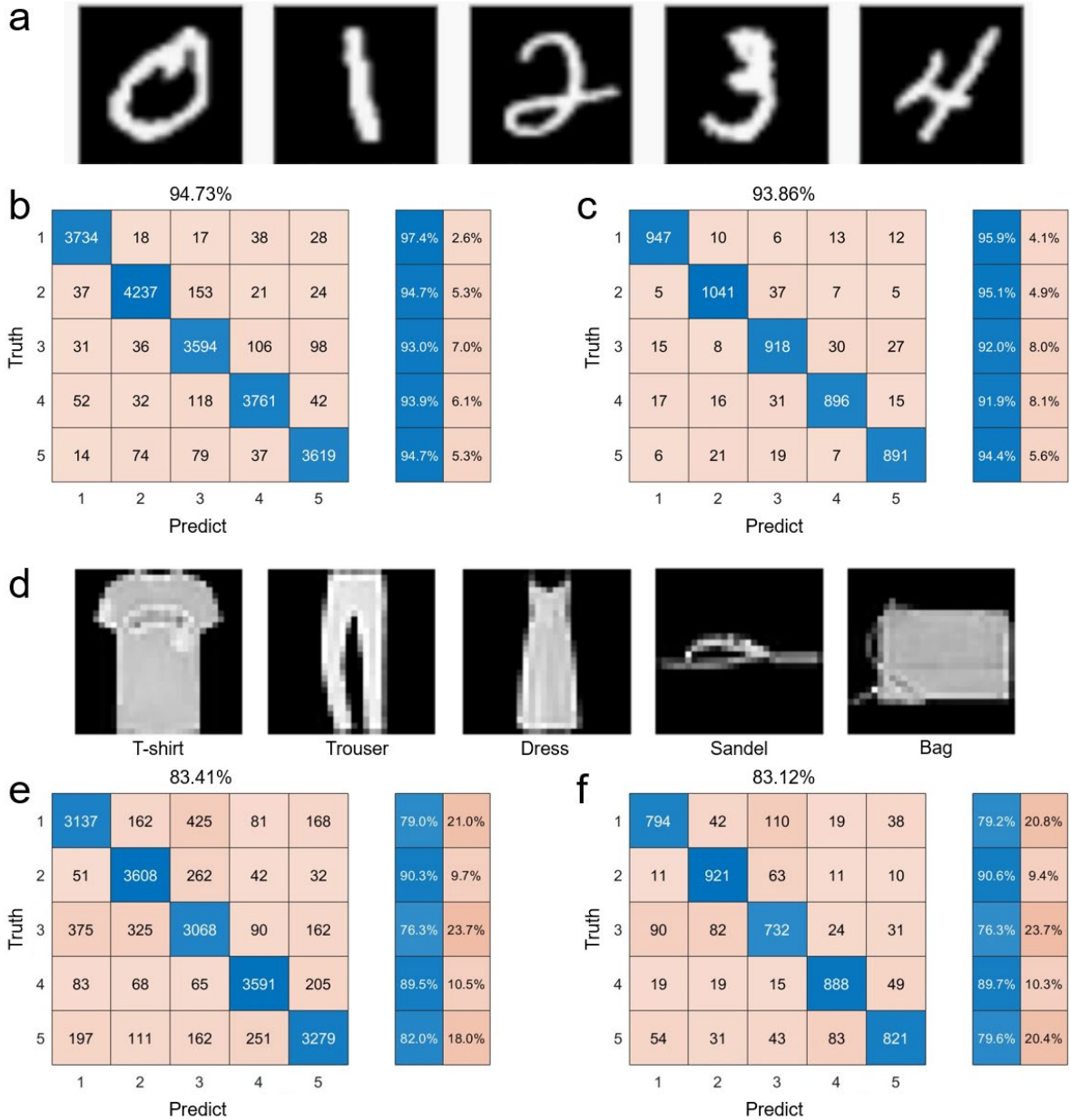

**Fig. S2 Five-category classification test results for typical images. a,** Schematic diagram of five-category MNIST handwritten digit classification examples. **b,** Classification results for the training set of five- category MNIST handwritten digits, with an accuracy of 94.73%. **c,** Classification results for the test set of five-category MNIST handwritten digits, with an accuracy of 93.86%. **d,** Schematic diagram of five- category Fashion-MNIST handwritten digit classification examples. **e,** Classification results for the training set of five-category Fashion-MNIST handwritten digits, with an accuracy of 83.41%. **f,** Classification results for the test set of five-category Fashion-MNIST handwritten digits, with an accuracy of 83.12%.

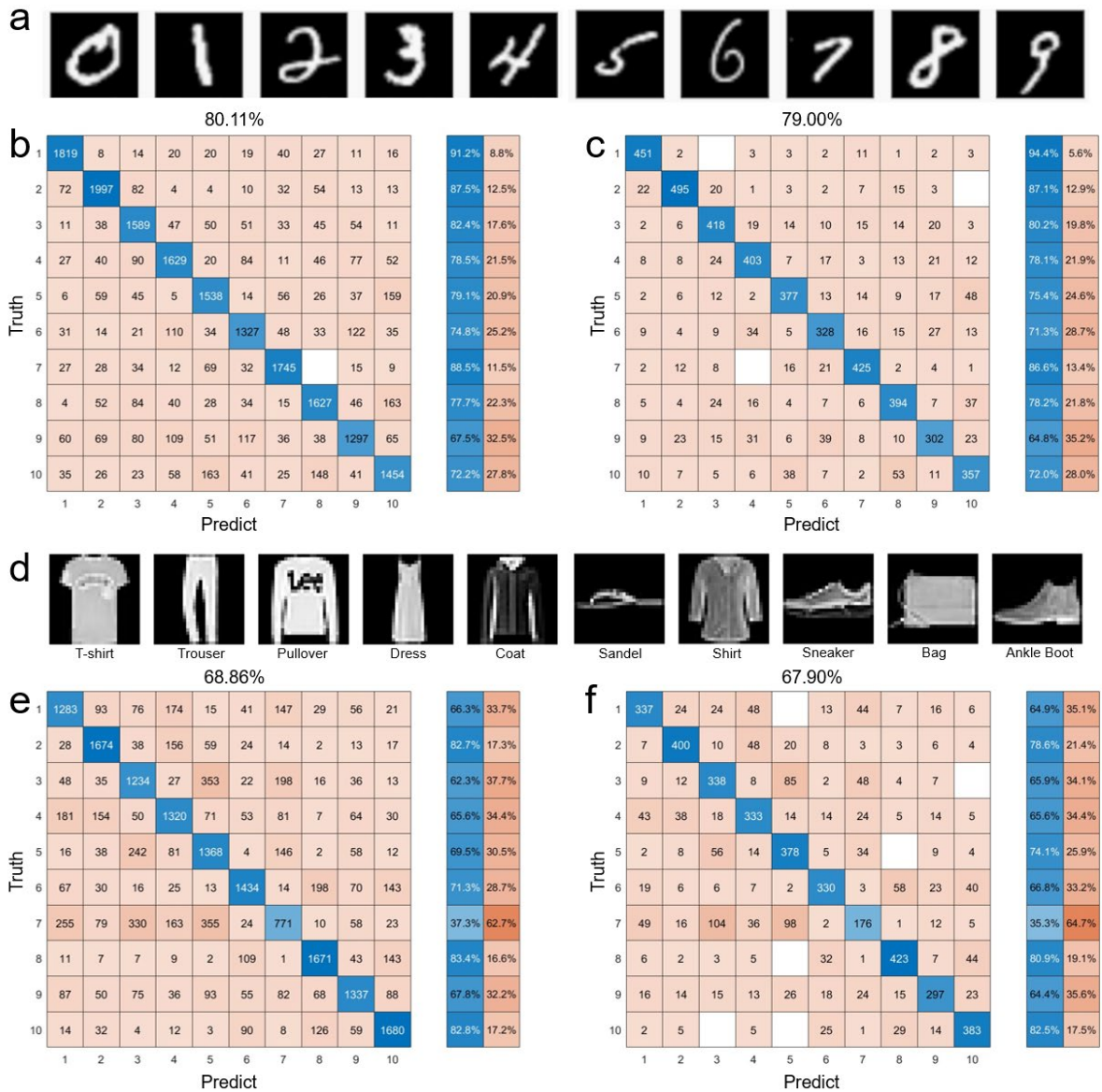

**Fig. S3 (Unmodified accuracy) Ten-category test results for typical images. a,** Schematic diagram of the ten-category MNIST handwritten digit classification sample. **b,** Classification results for the training set of the ten-category MNIST handwritten digits, with an accuracy of 80.11%. **c,** Classification results for the test set of the ten-category MNIST handwritten digits, with an accuracy of 79%. **d,** Schematic diagram of the ten-category Fashion-MNIST handwritten digit classification sample. **e,** Classification results for the training set of the ten-category Fashion-MNIST handwritten digits, with an accuracy of 68.86%. **f,** Classification results for the test set of the ten-category Fashion-MNIST handwritten digits, with an accuracy of 67.90%.

## Supplementary Note 2: Detailed structure of SPNN

In order to realize the programmable fully connected SPNN, the fundamental structures need to be designed. A fundamental SSPP unit is shown in Fig. S4a, in which the periodic length $p = 0.8$ mm, the height $h = 2$ mm, $a = 0.3$ mm, and $d = 0.5$ mm. In this case, the corresponding dispersion behaviors are simulated by CST simulation software, as shown in Fig. S4b. As the frequency increases, the dispersion curve gradually moves away from the light line, indicating stronger field confinement of SSPPs. This trend is an advantage for the minimization design. Thus, the SSPP-based coupler is designed, and the detailed parameters are listed in Table S1. Compared to the traditional microstrip coupler, the SSPP-based coupler has a smaller size. As the fundamental unit of the diffractive layer, this SSPP coupler exhibits consistent transmission and coupling parameters at the operating frequency, along with low reflection and good isolation, as demonstrated in Fig. S4f, which ensures unidirectional surface-wave propagation inside the network.

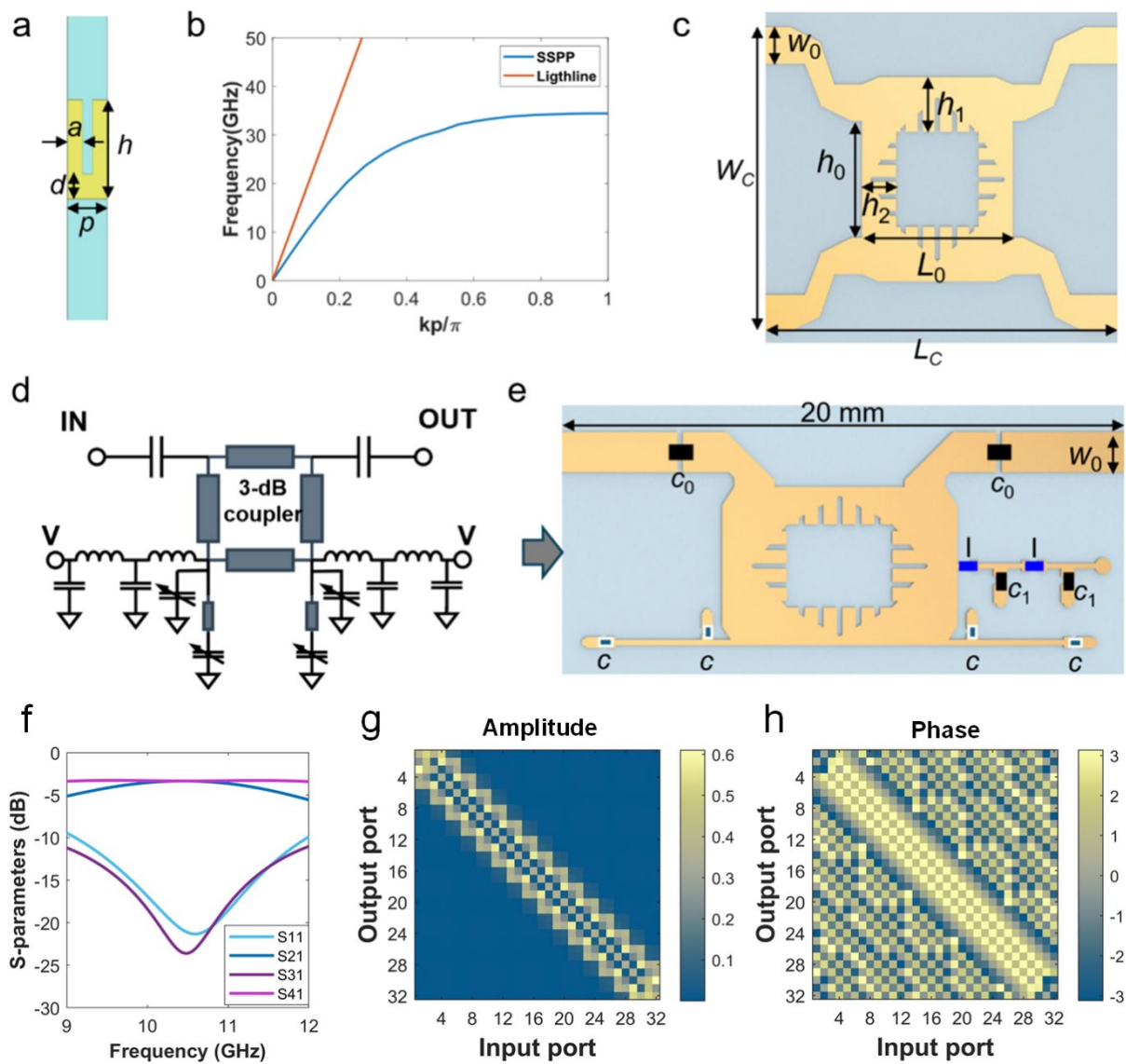

**Fig. S4 The fundamental structures of SPNN. a** and **b,** The SSPP unit and its corresponding dispersion characteristics. **c,** The SSPP coupler structure. **d,** The circuit diagram realizing the continuing phase range. **e,** The SSPP-based programmable phase shifter. f, Transmission and reflection coefficients of the coupler unit.

g, Amplitude transmission matrix of the 32-port coupling layer. h, Phase transmission matrix of the 32-port coupling layer.

**Table S1** Structure parameters

| $L_C$ | $W_C$ | $L_0$ | $H_0$ | $H_1$ | $H_2$ | $W_0$ | $C_1$ | $I$ | $C_0$ |
|---|---|---|---|---|---|---|---|---|---|
| 13 mm | 12.5 mm | 6.15 mm | 4.8 mm | 1.4 mm | 2.25 mm | 1.5 mm | 5.8 pF | 6.9 nH | 1 nF |

In addition, to realize a programmable phase-modulated layer, a 360° phase shifting range can be realized by using the reflection-type phase shifter[1]. This phase shifter consists of a 3-dB coupler and a reflective load with varactors (MAVR-011020-1411), as illustrated in Fig. S4d. Thus, the size of the phase shifter mainly depends on that of the coupler. For reducing the size of the phase shifter, we use the SSPP-based coupler shown in Fig. S4c. To control the bias voltage applied to the varactors, the designed feed circuit consists of the inductors and capacitors, with their values listed in Table 1. The reflection coefficient at the reflection-loaded ports is expressed by

$$\Gamma = \frac{j\left(-Z_1 + Z_1^2 C_v \omega \tan\theta\right) - Z_T \left(Z_1 C_v \omega + \tan\theta\right)}{j\left(-Z_1 + Z_1^2 C_v \omega \tan\theta\right) + Z_T \left(Z_1 C_v \omega + \tan\theta\right)} \quad (S1)$$

where the $Z_1$ and $\theta$ are the characteristic impedance and electric length of the transmission line integrated varactors, respectively. $Z_T$ and $C_v$ represent the port impedance of the SSPP coupler and the capacitance of the varactor, respectively. According to Eq. (S1), the corresponding phase of the phase shifter can be written as

$$\Delta\varphi = 2\left[\arctan\left(\frac{j}{Z_T}\left(\frac{-Z_1 + Z_1^2 C_v \omega \tan\theta}{Z_1 C_v \omega + \tan\theta}\right)_{max}\right) - \arctan\left(\frac{j}{Z_T}\left(\frac{-Z_1 + Z_1^2 C_v \omega \tan\theta}{Z_1 C_v \omega + \tan\theta}\right)_{min}\right)\right] \quad (S2)$$

For the selected varactor, the $C_v$ value varies from 0.025 to 0.25 pF when the applied bias voltage changes from 15 to 0V. Therefore, the maximum phase difference can be obtained according to Eq. (S2). Notably, the operating frequency mainly depends on the SSPP coupler. Simulated results show that when the capacitance ranges from 0.025 to 0.25 pF, the SSPP phase shifter achieves a 330° phase variation, which is sufficient for constructing the phase-modulated layer of the programmable SPNN. The amplitude and phase parameters of the constructed 32-input 32-output diffractive layer at 10.8 GHz are shown in Figs. S4g and S4h, respectively, where the relationship between the input and output signals can be freely controlled by cascading the phase-modulated layer.

# Supplementary Note 3: Training and test of dynamic-gesture classification

For the dynamic gesture recognition, the receiving signals (from the antenna array or from the output ports from the SPNN) could be regarded as a time-varying envelope that complies with the hand's movement. In the experiment, we trained the SPNN together with the digital linear classifier in silico based on the collected dataset. The sequentially collected dataset is grouped into the folded time dimension with a time interval of 1s. The detailed procedure is illustrated in Fig. S6c. The 16 individual antennas in the array (Fig. S5a) are connected to the 9th to 24th ports among the 32 input ports of the SPNN. The other ports are connected with matched loads and denoted by zeros in the analytical model. As illustrated in Figs. S5b, the target diffraction information as the training data is directly collected by the antenna array. Echo training data from different frequencies (f1-f3), beam scanning angles (b1-bn), antenna ports (C1-Cn), and time instants (t1-tn) form a multidimensional dataset containing the spatiotemporal information. For each training sample of the network, the input dimension is $32 \times 3 \times 2$. The three dimensions represent the number of input ports of SPNN, three discrete frequencies and the front/rear folded time slots respectively. In the forward model, we assume that the network is non-dispersive, i.e., the SPNN demonstrates a homogeneous transfer function for different frequencies. We select the 6th, 14th and 22nd at the output layer as the detection nodes. The detectors could perform the linear combination of different frequency components, which are designed as non-correlated during the injection. Therefore, the linear classifier receives the signal intensities components with the dimension of $3(\text{Ports}) \times 2(\text{Time Points})$ and performs the linear transformation to the final decision outputs. The coefficient matrix of the linear transformation has a size of $5 \times 6$. The first dimension corresponds to the designated five categories (four gestures and one void class).

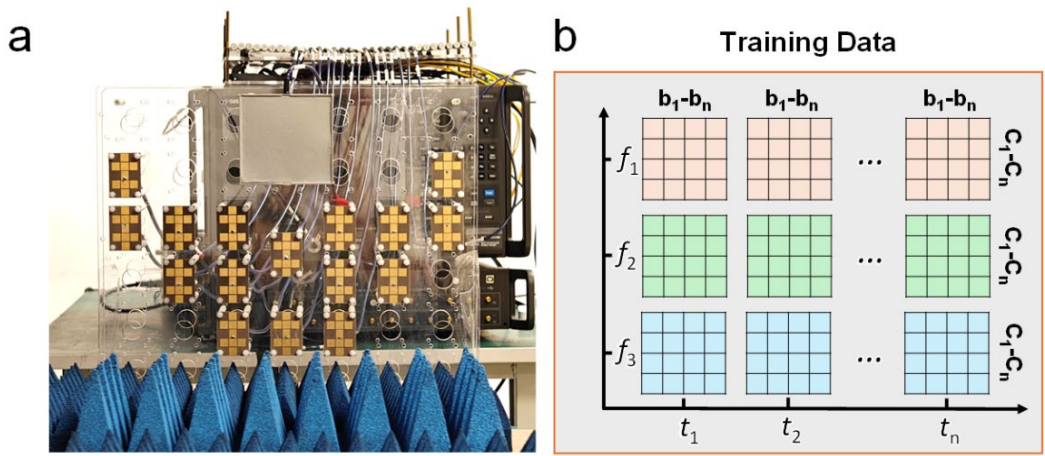

**Fig. S5 The training and test system illustration of dynamic-gesture classification. a,** Schematic diagram of the P-SPNN front-end transmitting (green frame) and receiving (blue frame) antenna arrays. **b,** Multi-frequency and multi-dimensional training data are constructed with the echo signals at three frequencies collected by the front-end antenna array.

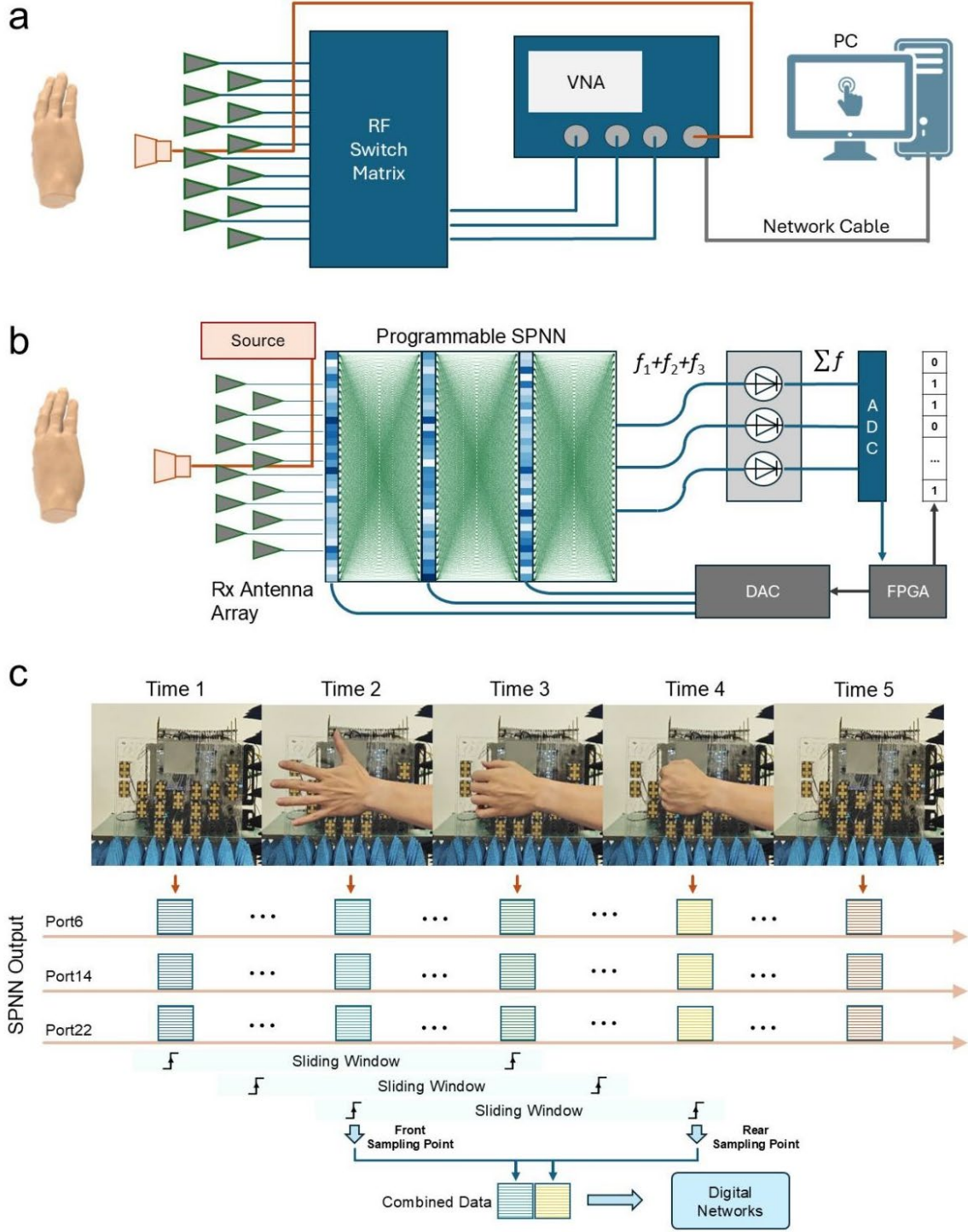

**Fig. S6 The training and test system illustration of dynamic-gesture classification. a,** The training data collection system of dynamic-gesture classification. **b,** The test system of dynamic-gesture classification. **c,** The time-domain data-processing illustration of the dynamic-gesture classification.

The analytical model of the network could be depicted by

$$Y_{SPNN}(:,i,j) = f_{SPNN}(X(:,i,j)) \qquad i = 1,2,3; j = 1,2 \qquad (S3)$$

$$X_{Linear} = flatten(sum(\|Y_{SPNN}\|^2, iDim = 2)) \tag{S4}$$

$$Y = WX_{Linear} + b \tag{S5}$$

Equation (S3) depicts the perception compressing inside the SPNN. During the propagation inside the waveguide structures, the network converts the information from the 16 input nodes into 3 nodes. The forward function $f_{SPNN}$ is constructed by cascading the fixed propagation layer and tunable phase-shifting layer. Based on the scale of the fabricated SPNN, three propagation layers and three modulation layers are adopted here. Equation (S4) depicts the energy conversions at the detectors, as well as the linear combination across different frequencies. Finally, the linear classifier is modeled by the simple linear transformation depicted in Eq. (S5). Each introduced layer is modeled via the Machine Learning Toolbox in MATLAB. In specific, the complex values in Eq. (S3)~(S4) are decomposed into two channels representing the real part and imaginary part respectively. We trained the network based on the auto-differential mechanism embedded in MATLAB. The optimizer is chosen as stochastic gradient descent with momentum (SGDM) with an initial learning rate of 0.01 and a minibatch size of 64.

In the experiment, we collected the samples from four different persons. The recipient was asked to perform each hand gesture combination 50 times. The sequential S-matrix data were collected from an RF switch matrix and a vector network analyzer, as shown in Fig. S6a. During the measurement, time stamps were recorded correspondingly, which were later used to synchronize the time axis for dynamic detection. For each measurement, a total number of 10 valid samples could be extracted for one of the four valid categories, while the other time-folded samples are all marked as the void class. Therefore, the training data has a size of 4*50*5*10 = 10000. We randomly select a portion of the void samples, so that the numbers for each category of samples are identical. Similar to the discussed procedures, we collected the validation data, which were separated from the training data during the network training.

In Fig. S7a-d, we present some visualized intermediate results inside the P-SPNN classification. The scatter plots show the intensity of the P-SPNN outputs, whose colors correspond to the feature distributions of the static gestures in the dynamic combination. From the divergence tendency of the point clouds, we could infer that the processed datasets after the SPNN possess are distinctly different in the limited feature space, which further facilitate the final linear layer inside the overall network. In accordance with the temporal shift, the prediction outputs (the normalized softmax output) for the four sets of dynamic gestures. The colors 1-4 represent the corresponding categories, with the green curve "none" serving as a no-gesture comparison. As shown in the figures, all gestures are accurately recognized within the fixed observing window. Although the recognition curves of Fig. S7b have some fluctuations, the value for "gesture 2" remains significantly higher than the other gestures.

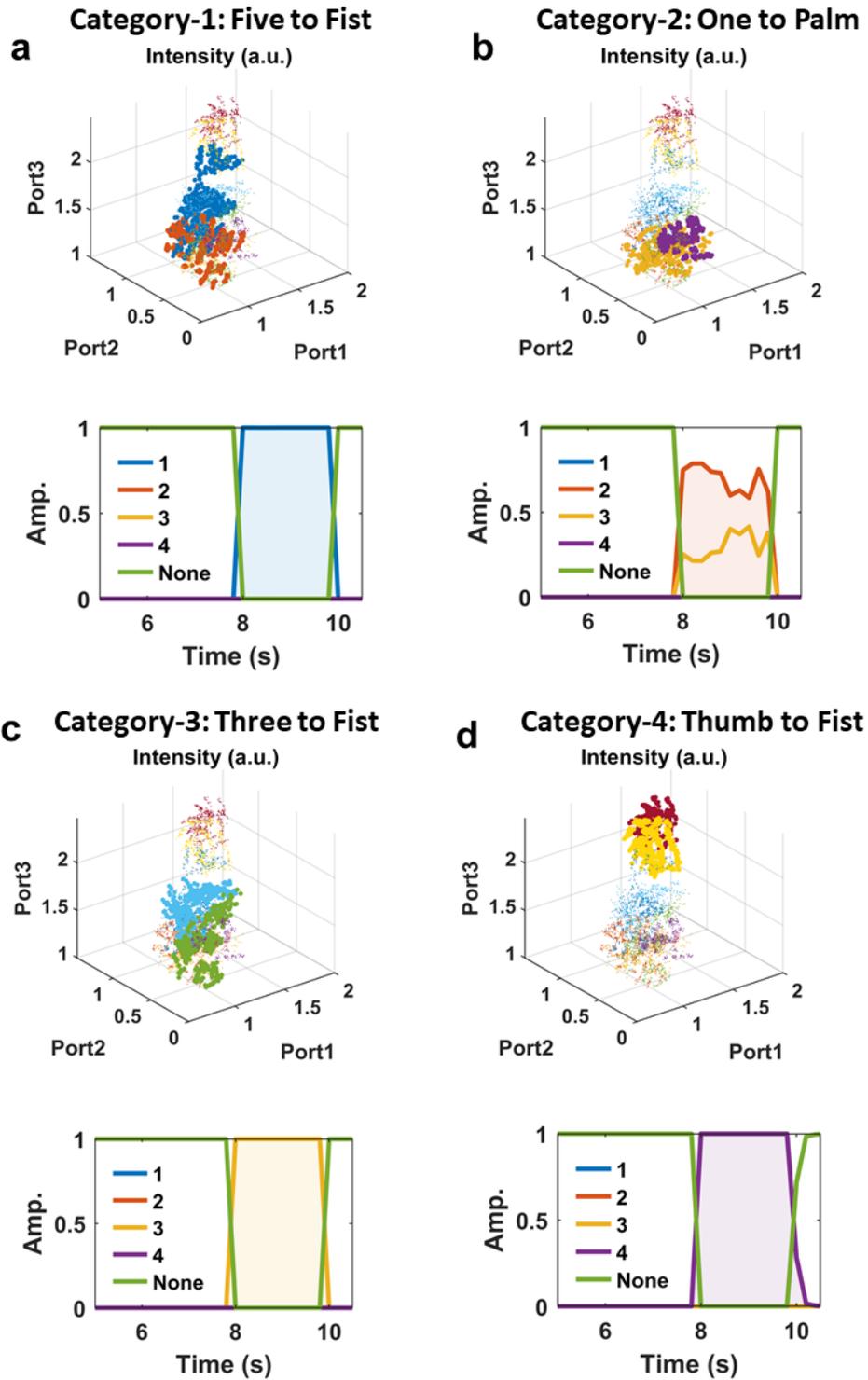

**Fig. S7 Some visualized intermediate results inside the P-SPNN classification.** The scatter plots and prediction outputs (the normalized softmax output) for four sets of dynamic gestures. **a,** Five to Fist. **b,** One to Palm. **c,** Three to Fist. **d,** Thumb to Fist.

# Supplementary Note 4: Description on the classification of person and car

In pedestrian and vehicle recognition applications, we designed the following combinations of recognition area scenarios for open roads where vehicles drive on the right. Figures S8a-c respectively show the pedestrian and vehicle recognition situation in the left, middle, and right areas directly in front of the vehicle from the perspective of a vehicle driving on the right, corresponding to different beam scanning angles. For the left area, as shown in Fig. S8a, we mainly consider whether there are oncoming vehicles within the dashed box area, which corresponds to the beam scanning angle range of -50° to -20°. For the central area, considering that both pedestrians and vehicles may appear directly in front, we set three scenarios, as shown in Fig. S8b. Figure S8c shows the target perception situation on the right side of the vehicle. Taking a two-way single-lane road as an example, the right side is mainly the pedestrian walkway area, where pedestrians may be waiting to cross the road. Therefore, the beam scanning range (20° to 50°) corresponding to this area includes whether there are people or not.

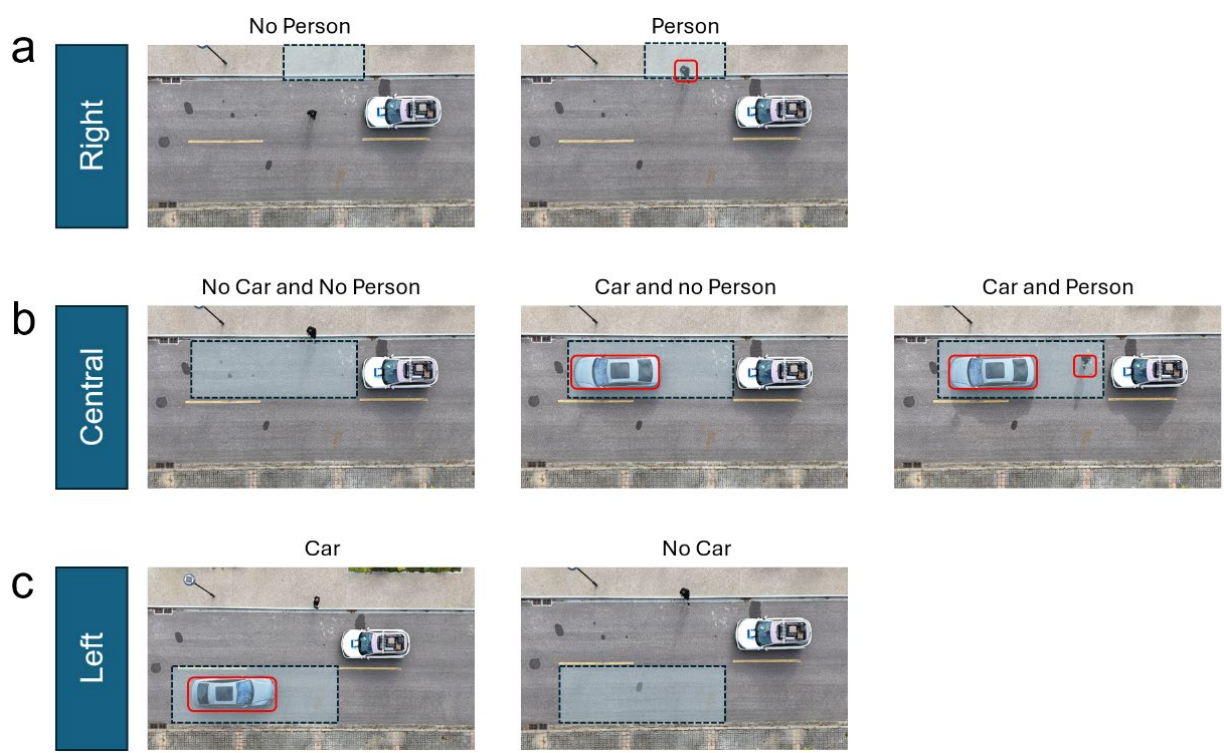

**Fig. S8 Illustration of classification scenarios in a human-vehicle recognition application. a,** Identification of whether a pedestrian is present on the right sidewalk. **b,** Classification of whether a person or vehicle is present in the lane directly ahead. **c,** Classification of whether a vehicle is present in the left oncoming lane.

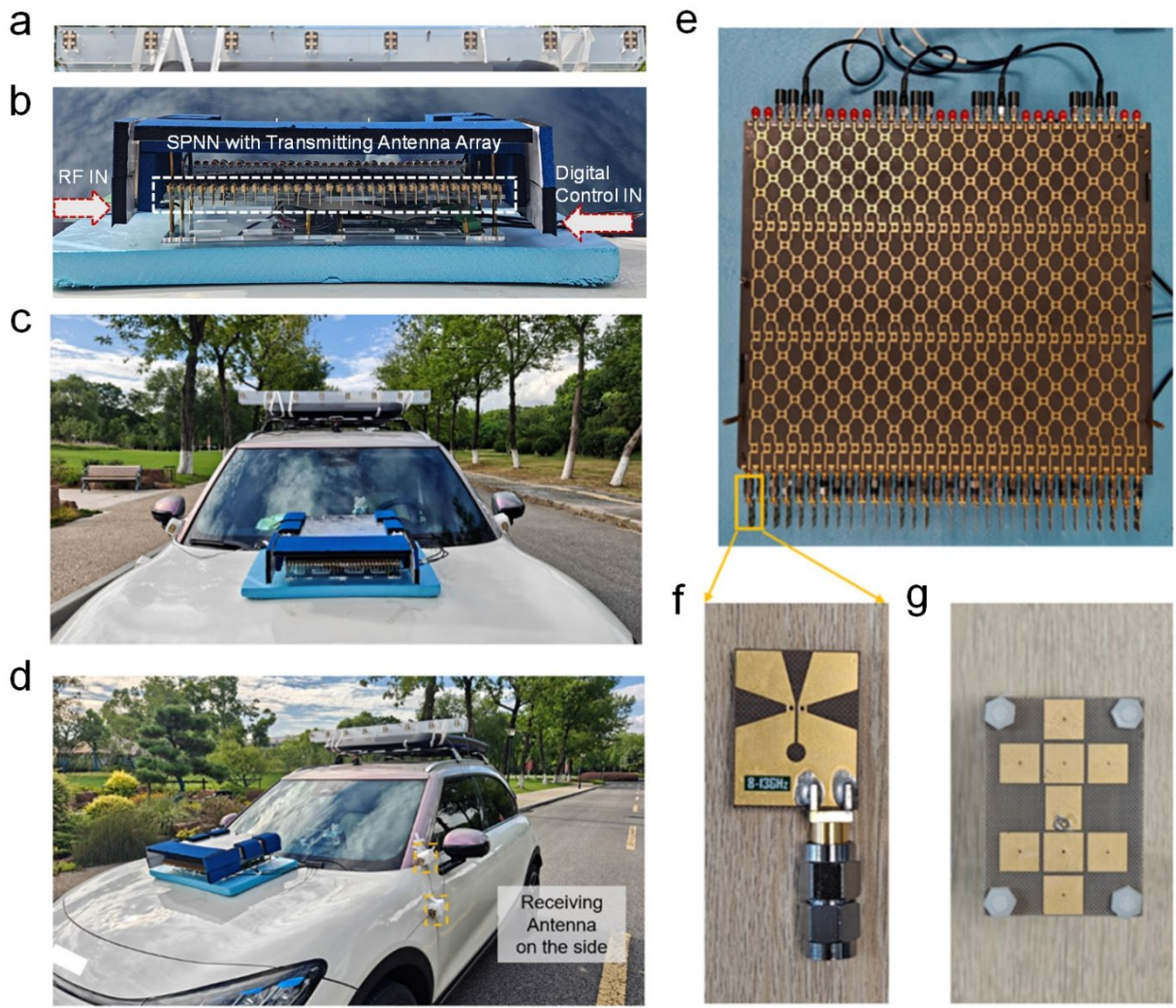

**Fig. S9 Installation details of the vehicle-mounted platform for person-vehicle recognition. a,** Eight antenna brackets on the vehicle roof for signal reception. **b,** Front view of the Tx SPNN. The programmable SPNN front end is connected to 32 antennas to form an antenna array to achieve beam scanning. The left and right sides are connected to the RF input and digital control lines, respectively. The Tx SPNN is coated with absorbing material to reduce the electromagnetic crosstalk. **c,** Front view of the vehicle-mounted platform. The transmitter SPNN is placed on the front of the hood to reduce interference from the transmit beam on the receiver. **d,** Side view of the vehicle-mounted platform. The two antennas on the vehicle side are located directly below the rearview mirror. **e,** Enlarged schematic diagram of the transmitter SPNN. The four input ports are excited by a one-to-four power splitter. **f,** Actual image of the transmitting broadband antenna. **g,** Actual image of the receiving microstrip antenna.

## Supplementary Note 5: Installation details of the vehicle-mounted platform for person-car recognition

For the person-car recognition test, we conducted experiments on open roads within the campus. We constructed the SPNN sensing platform based on a small Sport Utility Vehicle (SUV), as shown in Fig. S9. Figure S9a shows the receiving antenna array, which is fixed to an acrylic bracket for easy installation on the vehicle's roof. In actual testing, an absorbing shell (Fig. S9b) was used to reduce the backscattering interference on the receiving antenna. Figure S9c shows a front view of the vehicle-mounted experimental platform. In addition to the eight receiving antennas on the roof, there are two antennas on each side of the car. Considering that people and vehicles generate different scattering signals from different orientations, the multi-antenna layout on the roof and sides of the vehicle effectively receives scattered signals from different orientations. Figure S9d shows a side view of the vehicle-mounted experimental platform. The two side antennas are installed below the rearview mirrors, effectively receiving scattered echoes from people and vehicles on the left and right sides. Figure S9e shows a physical image of the Tx SPNN. The four input ports are connected to a power splitter via coaxial cables for simultaneous excitation. The output of the SPNN is directly connected to 32 antennas forming the antenna array. The transmitting antenna unit is shown in Fig. S9f, which adopts a Vivaldi-like wide antenna structure. The receiving antenna is shown in Fig. S9g, which adopts a microstrip antenna structure.

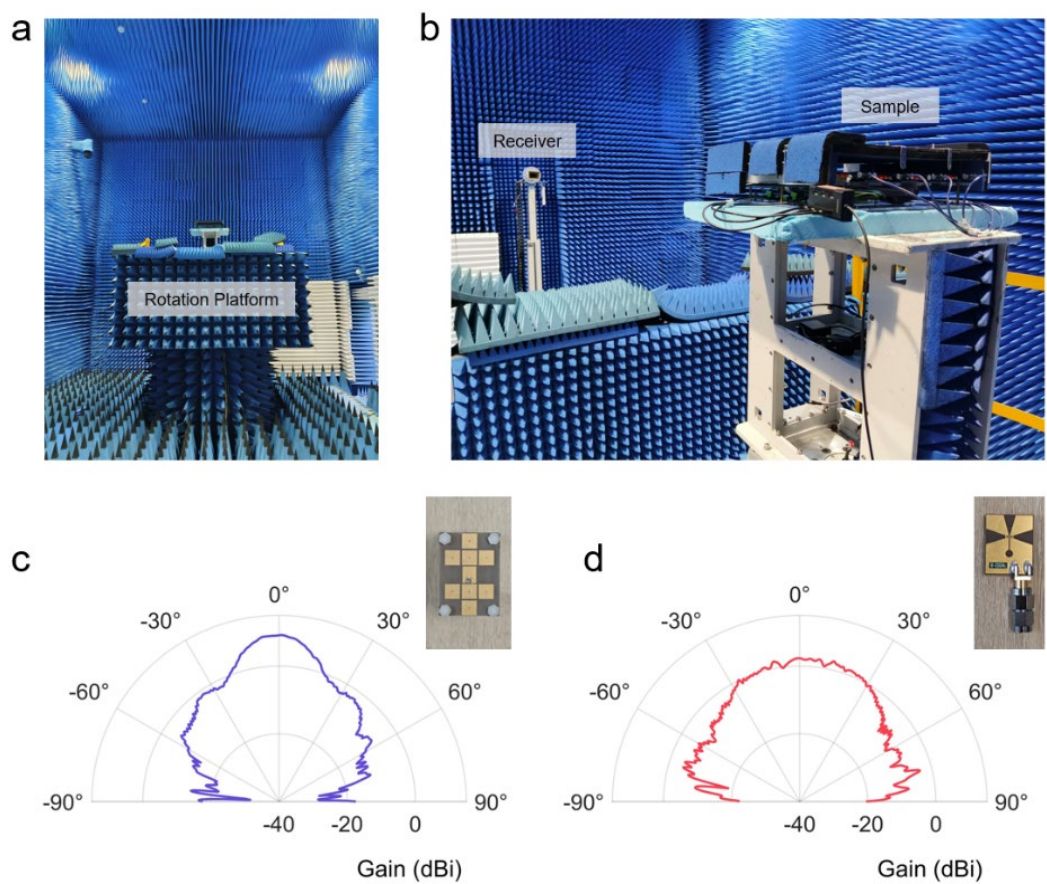

**Fig. S10 SPNN far-field experimental test in a standard microwave anechoic chamber. a,** Front view of the far-field test platform, where the test specimen is mounted on a rotating platform. **b,** Side view of the far-field test platform. **c,** The measured far-field result of the receiving antenna. **d,** The measured far-field result of the transmitting antenna.

## Supplementary Note 6: Far-field measurements of Tx SPNN for beam scanning

To ensure the far-field beam performance of the SPNN, we tested the beams of different scanning configurations of the Tx SPNN in a standard microwave anechoic chamber. The test scenario is shown in Fig. S10, where the Tx SPNN is mounted on a turntable, with the receiving antenna on the other side of the chamber (Fig. S10a). The turntable rotated in a two-dimensional plane to produce the planar far-field results, as shown in Fig. S10b. During the test, the Tx SPNN network weights remained constant, switching between weights for each different beam scanning angle. As per design requirements, a total of 11 beam scanning angles, or 11 sets of network weights, were tested. In addition, to ensure the performance of the transmit and receive antennas, we also tested their far-field patterns. At a center frequency of 10.6 GHz, the transmit and receive antenna patterns are shown in Fig. 8c,d. Based on the measured results, the individual transmitting and receiving antenna gains are 2.6dBi, 9.2dBi, respectively.

## Supplementary Note 7: Training of person-car classification

In this section, we would show the detailed algorithm framework for person-car classification task. Figure 5c in the manuscript demonstrates the algorithm diagram. Compared with the algorithm used in hand-gesture recognition, the new algorithm introduces two major updates. Firstly, an addition SPNN is used to generate dynamic apertures, which increases the detection resolution and expands the perception dimension inside the spatial domain. Secondly, the algorithm adopts a two-stage training process, i.e. pre-training and fine-training, which helps to network to be resilient to environment variation. In the pre-training, all parameters in the built neural network model are trained based on the collected dataset. Then, in the real deployment, we would collect the analog part computing results output and tune the few digital parameters in the linear classifier accordingly while retaining the tuning parameters in the analog hardware. We term this post process fine-training, as its training cost is fairly small compared with the pre-training process. However, it would significantly increase the network's performance and counters the environment changes in real scenarios.

Similar to the previous hand-gesture recognition task, we use the frequency-dimension to introduce the perception ability for the distance measurement. Around the centering frequency of 10.5GHz, we select two additional frequency points, 10.45GHz and 10.55GHz. Again, we assume that the neural network would demonstrate a quasi-non-dispersive behavior and only the free-propagation process would introduce different phase-delay for three frequencies. In the experiment, one of the SPNN platforms (Tx SPNN) is used for beam scanning. The phase codings for each pattern are generated using the cross-entropy loss function defined on the far-field results.

$$y_{out} = \prod \Phi_i W_i x_0 \tag{S6}$$

$$loss = \log\left(sum(W_p y_{out} \cdot F_o)\right) \tag{S7}$$

$x_0$ represents the input vector of the neural network, which is a zero-element vector except the 13th,15th,17th, and 19th elements being ones. The non-zero elements correspond to the four injected ports connected to one equally-split 4-port power divider. $F_o$ is the one-hot-coding vector for the designated 11 beam directions (-50° to +50°). The propagation matrix $W_p$ is defined by the antenna array theory. The angular resolution of $F_0$ and $W_p$ is 1 degree. For each beam direction, the training process is carried out once based on SGD algorithm. The acquired codings are converted into the desired voltages applied to the varactors and loaded inside the controller. At each triggering signal, the controller periodically refreshes the codings.

In the real-world scenario, we have divided the potential obstacle position into three regions: left (-50°~-20°), middle(-10°~10°) and right (20°~50°). For each region area, we would train an independent phase coding and linear classifier for potential obstacle recognition. Since the beam scanning at the first SPNN platform inheritably requires the coding switching, the three sets of classification tasks at the second SPNN would not introduce additional processing delay. Considering the habit of motor vehicles and traffic rules, we assume that only vehicles and pedestrians would exist in the left and right regions respectively, while both could exist in the middle region. Therefore, including the void (empty) class, three regions correspond to a two-class, three-class, two-class classification task. Based on the collected samples (See Methods), we trained the network using the Machine Learning Toolbox on the MATLAB Platform. Similar to the hand gesture classification task, the network contains three phase-modulation layers and one linear classifier for optimization.

In the fine-training process, we collected the raw Rx SPNN signals in the open-world experiments and re-train the linear classifier at the last layer of the network. Such procedure requires a relative small sample size with a negligible re-training resources consumption, as the classifier itself has a very limited training variables.

## Supplementary Note 8: Processing time of P-SPNN

We provide a detailed description of the processing time of P-SPNN. Its single perception time is primarily determined by the time-of-flight of electromagnetic waves, the propagation time within the diffractive network, and the processing time on the digital backend. As shown in Fig. S11a, the electromagnetic wave signal is first emitted by the Tx SPNN, reflected by the target, then enters the Rx SPNN, and directly produces the computation result (in the form of intensity) at the output of the Rx SPNN. This result can be read by an ADC and converted into information suitable for digital processing. In the human-vehicle recognition perception experiment, the beam switching of the Tx SPNN and the ADC acquisition are controlled by a synchronization signal; each beam switch triggers one data acquisition from the Rx SPNN. As illustrated in Fig. S11b, each test computation includes the time for 11 beam switches and the processing time of the MCU digital backend. The detection time per beam angle consists of three parts: (1) the time-of-flight of the electromagnetic wave and the response time of the detector (approximately 25 ns altogether); (2) the readout time of the ADC for the output voltage of the detector; and (3) the network weight switching time of the Tx SPNN, mainly contributed by the switching of the DAC matrix, which is about 8.95 µs. Therefore, the total time for 11 beam switches and P-SPNN computation is 103.24 µs, with the majority of this time (about 98 µs) originating from the high-precision DAC switching used to control the varactors in the programmable neurons. It should be noted that this portion of time can be significantly reduced by employing relatively lower-precision but higher-speed DAC chips. In Fig. S11c, we further present the oscilloscope test results of the DAC voltage switching speed, where the switching time for 11 voltage sets is 98.4 µs. The four channels refer to the four control channels under the same DAC chip. It can be observed that under this switching speed, different switching voltage states remain stable. To facilitate data capture by the oscilloscope, the control test employed a method of cyclically switching staircase voltages with the DAC, analogous to the switching of different beam network weights in practice.

## Supplementary Note 9: Experimental test of the SPNN diffraction layer for S-parameter matrix

Considering the high resonant property of SPP structures, the EM characteristics of the adopted coupler and phase shifter modules are sensitive to various errors, including the pattern deviation caused by chemical etching in PCB fabrication, substrate dielectric fluctuation, imperfect impedance matching and cross coupling between adjacent ports. Although these errors do not fundamentally compromise the skeleton of the network, they would cause a discrepancy between the real platform and the numeric model for optimization. When the layer number increases, the errors would accumulate and damage the network's performance. To solve this, one possible method is to transfer the training process directly to the analog platform. However, due to the lack of efficient gradient-based algorithm realization, the converging speed during the optimization is usually compromised [2]. Here, we try to minimize the mentioned errors by using the measured data from a twin-fabricated sample. In the manufacturing of the SPNN boards, several network sub-modules are fabricated together, as shown in Fig. S12a. Since they use the same batch of substrate, pattern masks, we may infer that they share an identical deviation between the ideal model. Therefore, through this twin calibration method, we could acquire the real parametric data in actual samples. Based on the twin samples shown in Fig. S12a, we measured the 32-in 32-out propagation matrix (diffraction matrix), which is later adopted in the training process. Figure S12b and Figure S12c show the amplitude and phase distribution of the diffraction matrix at 10.8 GHz.

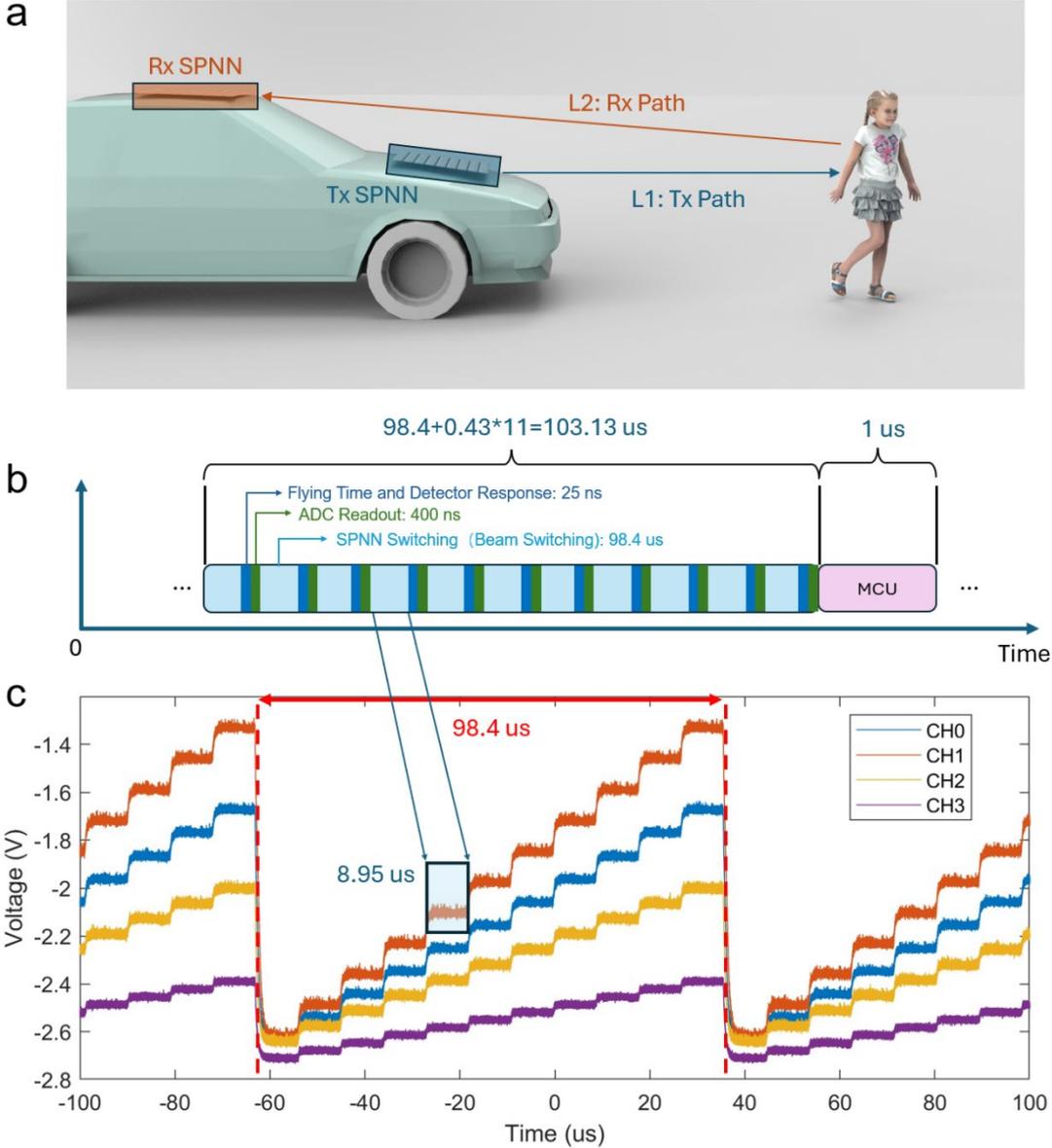

**Fig. S11 The illustration of the processing time of P-SPNN. a,** The schematic of the P-SPNN Processing time, including the flying time in air and diffraction time in SPNN. **b,** Time allocation instructions for each measurement group. **c,** Oscilloscope test results for DAC switching, where four channels are switched into different voltage states.

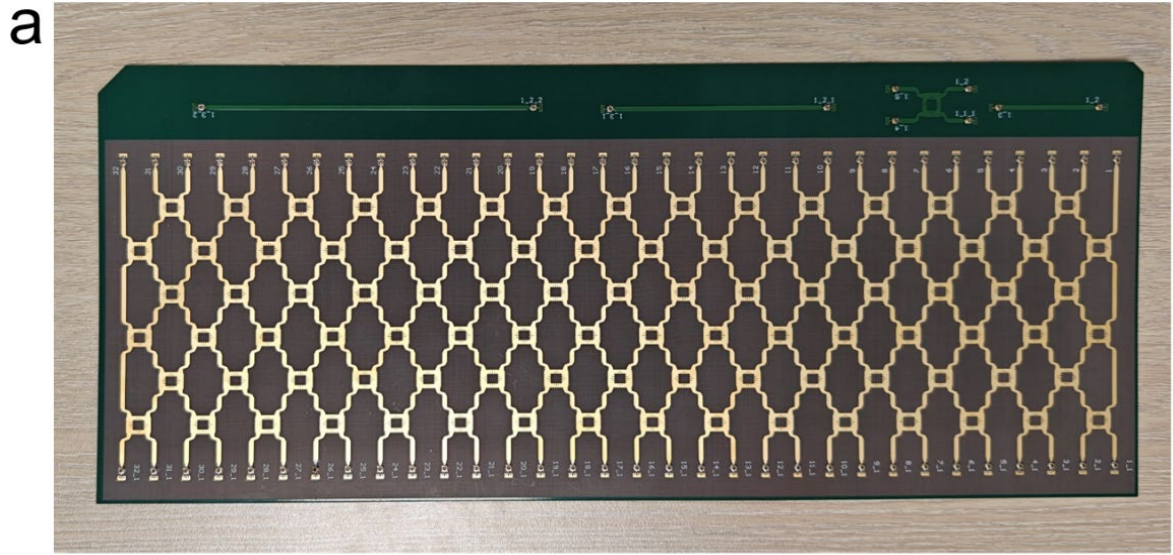

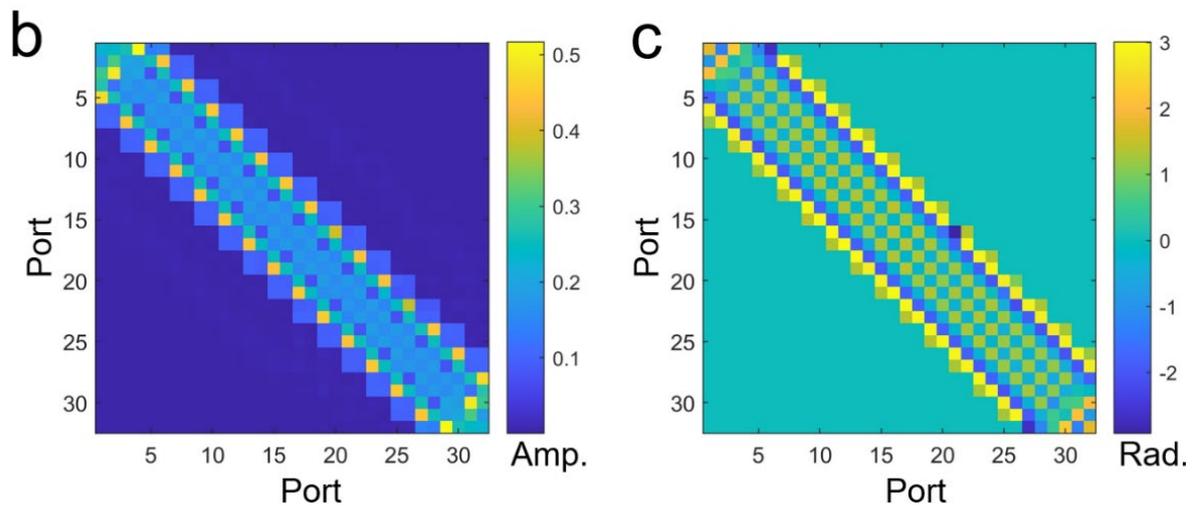

**Fig. S12 Experimental test of the SPNN diffraction layer for S-parameter matrix. a,** Fabricated diffraction matrix verification board. **b,** Measured amplitude matrix of the diffraction layer. **c,** Measured phase matrix of the diffraction layer.

## Supplementary Note 10: Scale expansion of P-SPNN

Here, we provide a possible scaling approach for the proposed P-SPNN platforms. As shown in Fig.S13, the two-dimension skeleton could be grouped by sewing different panels of current P-SPNN architecture. While the cascading along the propagation direction (from input to output) is fairly easy concatenation the ports of different sub-panel, the vertical expansion would need a modification of the basic module in the SPNN. As shown in the Fig.S14, the upper and bottom boundary of the waveguide structures in the P-SPNN are required to be dismantled so that the propagation of signals across vertical panels become available. In specific, the coupler port at the margin of the P-SPNN needs to be modified to an external SMA (yellow arrow), and the cascading with the front and rear couplers needs to be disconnected (blue cross). It could be proved that such configuration would not introduction the unwanted back-scattering inside the guided wave structurers and the original numeric model of propagation process would sustain in this proposed scaled P-SPNN architecture.

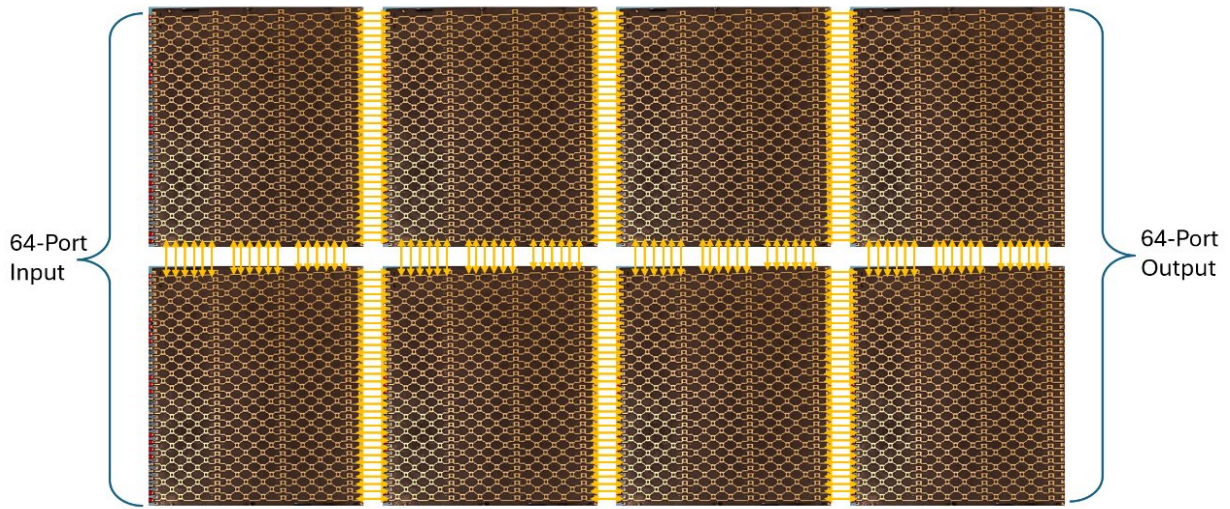

**Fig. S13 Scale Expansion of 8 P-SPNN.** Each P-SPNN is interconnected through adjacent SMA connectors, forming a 768-neuron network with 64 input and output ports.

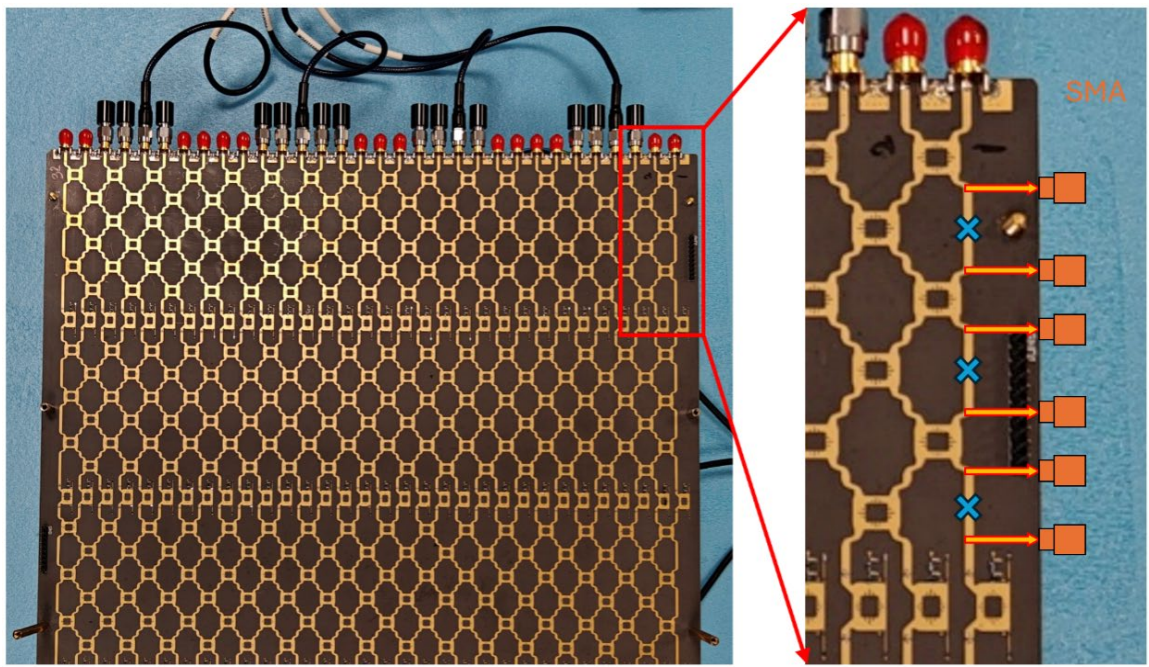

**Fig. S14 The port modification of the P-SPNN for scale expansion.** The side coupler port of the P-SPNN needs to be modified to an external SMA (yellow arrow), and the cascading with the front and rear couplers needs to be disconnected (blue cross).

# References

1  Burdin, F., Iskandar, Z., Podevin, F. & Ferrari, P. Design of Compact Reflection-Type Phase Shifters With High Figure-of-Merit. *IEEE Transactions on Microwave Theory and Techniques* **63**, 1883-1893, doi:10.1109/tmtt.2015.2428242 (2015).

2  Bandyopadhyay, S. *et al.* Single-chip photonic deep neural network with forward-only training. *Nature Photonics* **18**, 1335-1343, doi:10.1038/s41566-024-01567-z (2024).